\documentclass[prd,twocolumn,citeautoscript,superscriptaddress,notitlepage, longbibliography,bibnotes,footinbib]{revtex4-1}   
\pdfoutput=1  
\usepackage{amsmath,amssymb,mathrsfs,bm,feynmf,setspace,xspace,soul,empheq}  
\usepackage{graphicx}       
\usepackage[tight]{subfigure}          
\usepackage{color} 
\usepackage[colorlinks=true]{hyperref}      
\hypersetup{
    bookmarks=true,         
    unicode=false,          
    pdftoolbar=true,        
    pdfmenubar=true,        
    pdffitwindow=false,     
    pdfstartview={FitH},    
    pdftitle={My title},    
    pdfauthor={Author},     
    pdfsubject={Subject},   
    pdfcreator={Creator},   
    pdfproducer={Producer}, 
    pdfkeywords={keyword1} {key2} {key3}, 
    pdfnewwindow=true,      
    colorlinks=true,       
    linkcolor=magenta, 
    citecolor=blue,        
    filecolor=magenta,      
    urlcolor=cyan           
} 

\newcommand{\w}{\omega}

\newcommand{\De}{\Delta} 
\newcommand{\G}{\Gamma}

\newcommand{\beq}{\begin{equation}}
\newcommand{\eeq}{\end{equation}}
\newcommand{\ba}{\begin{array}{ccc}}
\newcommand{\ea}{\end{array}}
\newcommand{\nn}{\nonumber \\}

\def\bea{\begin{eqnarray}}
\def\eea{\end{eqnarray}}
\newcommand{\bml}{\begin{multline}}

\newcommand{\eeqm}{\end{multline}}

\newcommand{\bsp}{\begin{split}}
\newcommand{\esp}{\end{split}}

\renewcommand{\b}[1]{{\bf #1}}

\newcommand{\mc}{\mathcal}

\newcommand{\req}[1]{Eq.\thinspace(\ref{eq:#1})}

\newcommand{\ie}{{\em i.e.\/}\@\xspace} 
\newcommand{\eg}{{\em e.g.\/}\@\xspace} 
\newcommand{\ts}{\thinspace{}}

\DeclareMathOperator{\im}{Im}
\DeclareMathOperator{\re}{Re}

\newcommand{\bl}[1]{{\color{black}#1}} 

\newcommand{\mo}{\mc O} 
 
\newcommand{\gap}{m_0}

\begin{document} 

\title{Dynamical response near quantum critical points}   
\author{Andrew Lucas} 
\email{ajlucas@stanford.edu}
\affiliation{Department of Physics, Harvard University, Cambridge, Massachusetts, 02138, USA}
\affiliation{Department of Physics, Stanford University, Stanford, California, 94305, USA}

\author{Snir Gazit}
\affiliation{Department of Physics, University of California, Berkeley, Berkeley, California, 94720, USA}

\author{Daniel Podolsky}   
\affiliation{Physics Department, Technion, 32000 Haifa, Israel}

\author{William Witczak-Krempa}  
\email{w.witczak-krempa@umontreal.ca} 
\affiliation{Department of Physics, Harvard University, Cambridge, Massachusetts, 02138, USA}
\affiliation{D\'epartement de physique, Universit\'e de Montr\'eal, Montr\'eal (Qu\'ebec), H3C 3J7, Canada}

 \date{\today} 

\begin{abstract} 
We study high frequency response functions, notably the optical conductivity, 
in the vicinity of quantum critical points (QCPs) by allowing for both detuning from the critical coupling 
and finite temperature. We consider general dimensions and dynamical exponents.
This leads to a unified understanding of sum rules.
In systems with emergent Lorentz invariance, 
powerful methods from conformal
field theory allow us to fix the high frequency response in terms of universal coefficients.  
We test our predictions analytically 
in the large-$N$ O$(N)$ model and using the gauge-gravity duality, and numerically
via Quantum Monte Carlo simulations on a lattice model hosting the interacting 
superfluid-insulator QCP.   In superfluid phases, interacting Goldstone bosons qualitatively change 
the high frequency optical conductivity, and the corresponding sum rule. 
\end{abstract}
 
\maketitle 

\begin{figure}
  \center
  \includegraphics[scale=.52]{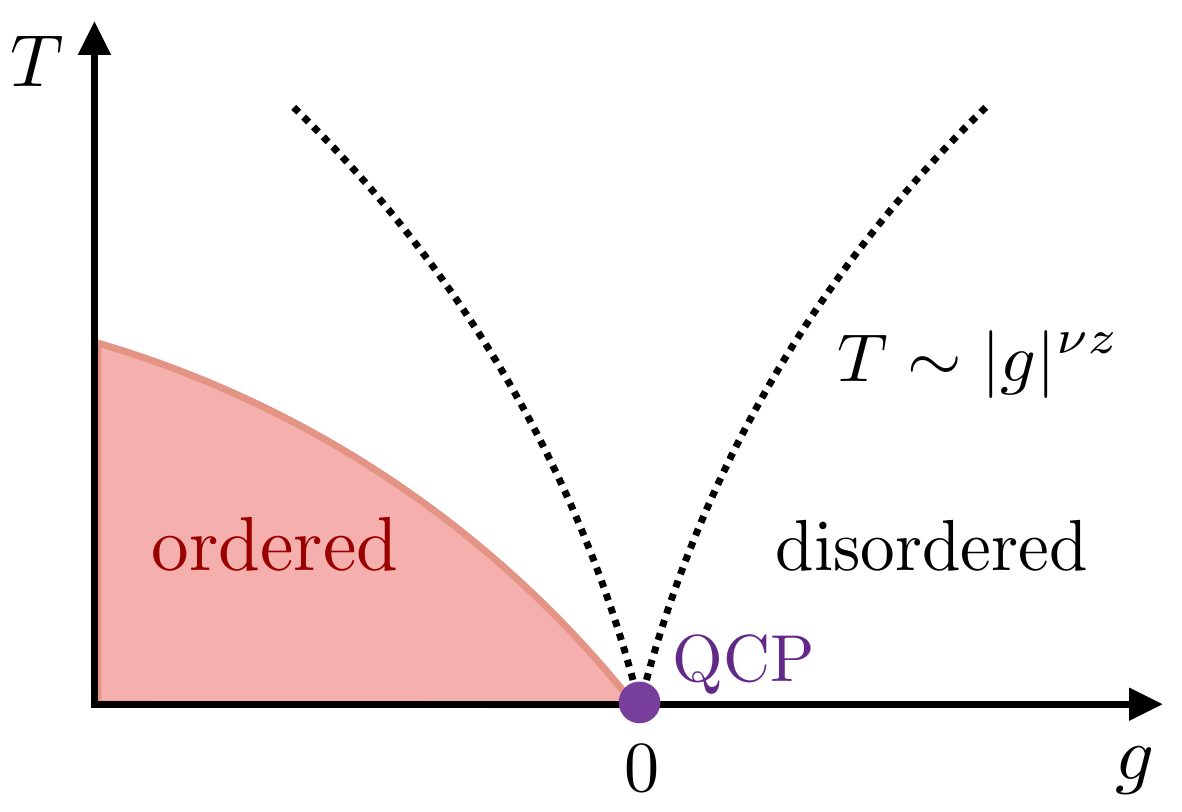} 
  \caption{Phase diagram near a canonical quantum critical point. $g$ is the non-thermal coupling
    that needs to be tuned. The dotted lines roughly delimit the quantum critical ``fan''. }  
  \label{fig:sigmap}  
  \centering   
\end{figure}

A quantum critical point (QCP) is a zero-temperature phase transition, driven by quantum fluctuations, 
reached by tuning a non-thermal parameter such as a magnetic field \cite{ssbook}, as shown in Fig.\thinspace\ref{fig:sigmap}.  
Proximity to a QCP alters many observables, even if the (detuned) ground state is otherwise conventional.  
Of particular importance are dynamical response functions such as the 
optical conductivity $\sigma(\omega)$ \cite{cha,damle1997,ssbook,smakov,will-mit,myers11,WS12,nat,Chen14,Gazit13,Katz,Gazit14,susy,Myers:2016wsu},        
where changing the frequency probes physics at different energy scales set by the non-thermal detuning and by the temperature.  
What often complicates the analysis of the real-time dynamics, especially on short time scales, 
is the destruction of quasiparticles at the QCP, and the corresponding abundance of incoherent excitations at finite but small detuning. 

\bl{In this letter, we focus on a large family of non-metallic QCPs \cite{ssbook} found in magnetic insulators,
Dirac semimetals, cold atomic gases in optical lattices \cite{Spielman07,Chin12,Endres12}, 
thin film superconductors or arrays of Josephson junctions \cite{cha}. This will serve as comparison ground
for the more intricate metallic QCPs occuring in heavy fermion materials for example \cite{Gegenwart08}.} 
Specifically, we study how the detuning of the non-thermal parameter from its critical value, 
as well as temperature, modify the optical conductivity.  
In particular, our analysis at large frequencies is not restricted to the quantum critical fan.    
We derive sum rules for the conductivity that generalize the standard $f$-sum rule \cite{Mahan}     
to the scaling regime near QCPs.  Our methods are not perturbative in any interaction strength. 
We test our predictions using large-scale quantum Monte Carlo simulations of
an interacting superfluid-insulator QCP.
While our focus is on the portions of the phase diagram smoothly connected to the critical fan, we also point out the qualitative changes to $\sigma(\omega)$ and the resulting sum rules which result from interacting Goldstone bosons in broken-symmetry phases.    

{\bf Setup:} Let us consider a system near a QCP that is reached by tuning a non-thermal parameter $g$ to zero. 
We work in the universal scaling regime, at frequencies smaller than microscopic (UV) scales, and 
assume that hyperscaling is obeyed.     
Such a system is described by the following low-energy action in $d$ spatial dimensions:
\begin{align} \label{eq:action}
  S= S_{\rm critical}  - g \int\! \mathrm d t\, \mathrm d^d \b x  \,\, \mo(t,\b x),
\end{align}
where $\mo$ is the only relevant operator whose coupling $g$ necessitates fine-tuning;
it has (spatial) scaling dimension 
\begin{align} \label{eq:Delta}
  \Delta=d+z-1/\nu, 
\end{align} 
where $\nu\!>\!0$ is the correlation length critical exponent, and $z$ is the dynamical exponent. 
The equal-time 2-point function of $\mo$ at the QCP is thus $\langle \mo(0,\b x) \mo(0,0)\rangle\propto 1/|\b x|^{2\Delta}$.
For example at the superfluid-insulator QCP in 2d belonging to the Wilson-Fisher universality class, 
$\mo\sim\phi_a\phi_a$ is the ``mass'' term
of the 2-component order parameter field $\phi_a$.  
At $T\!=\!0$, the correlation length diverges as  
$\xi\sim g^{-\nu}$ on the insulator side. 

We are interested in probing the properties of the nearly critical system  
by studying dynamical response functions such as the
optical conductivity: $\sigma(\omega)\!=\!\tfrac{1}{\mathrm{i}\omega}\langle J_x(-\omega)J_x(\omega)\rangle_{g,T}$, 
where $\omega$ is the frequency, and $J$ is the current operator that enters in the retarded correlator. 
Near the QCP,
the conductivity will obey scaling:
$  \sigma(\omega) = 
\omega^{(d-2)/z} f_\pm\! \left(\frac{\omega}{|g|^{z\nu}}, \frac{\omega}{T} \right)$,
where $f_\pm$ is a dimensionless scaling function that depends on which side of the transition the system is poised. 
\bl{We have set $\hbar=k_{\mathrm{B}}=1$, and the charge $Q\!=\!1$ and $c\!=\!1$, where $c$ appears in the energy scale $c |\b k|^z$.}
Other response functions such as order parameter susceptibilities or the shear viscosity will have an analogous structure.  

{\bf Large frequencies:}
In this letter we focus on the behavior of the conductivity at high frequencies 
$\omega\gg T,|g|^{z\nu}$, which allows us to controllably study the deviations away 
from criticality.  The resulting asymptotics will also serve as the key 
ingredient in the derivation of sum rules for the response functions. 
Our first main result is that the asymptotic behavior is 
\begin{multline} \label{eq:asympt}
  \sigma(\omega) = (\mathrm{i}\omega)^{(d-2)/z} \\ 
\bigg( \sigma_\infty 
+ c_1 \frac{g}{(\mathrm{i}\w)^{(d+z-\Delta)/z}} 
 + c_2 \frac{\langle \mo \rangle_{\!g,T}}{(\mathrm{i}\w)^{\Delta/z}} + \dotsb 
 \bigg),
\end{multline}
where $\sigma_\infty$, $c_{1,2}$ are real constants fixed by the universality class, independent of detuning and $T$. 
The $\sigma_\infty$ term is the conductivity of the critical theory; the $c_{1,2}$ terms arise from deviations from the QCP due to detuning and temperature. Note that the $c_1$ term in brackets simply scales as $\omega^{-1/\nu}$, by virtue of (\ref{eq:Delta}). 
In odd $d$, the imaginary part of $\sigma$ can have 
a non-universal logarithmic contribution, not written here. For simplicity, we consider the generic case where the $c_{1,2}$ 
power-laws are not equal, and more generally do not differ by $2n/z$ ($n$ being an integer), \ie   
$2\Delta\!\neq\! d+z+2n$.\footnote{When this condition is not satisfied, additional logarithms appear in Eq.~(\ref{eq:asympt}).}      


When $z\!=\! 1$, recent work has derived \cite{Katz} the $c_2$ term in Eq.~(\ref{eq:asympt})
at $T\!>\!0$ but zero detuning $g\!=\!0$.      
Here, we identify the new effects coming from detuning, and their interplay with temperature. 
In particular, the $c_1$ term purely arises from $g$ and can have important consequences on the dynamics.
Its existence was glimpsed deep in the quantum critical fan, $T\!\gg\!|g|^{z\nu}$, in a specific AdS/CFT calculation \cite{Myers:2016wsu}, and in fact holds much more broadly.
For CFTs ($z\!=\!1$) we will derive \req{asympt}, present a universal expression for $c_1/c_2$, and confirm our predictions with two independent 
computations in non-trivial CFTs.   For $z\!\neq\!1$, we provide a general scaling argument for the $c_1$ term, 
and confirm that Eq.~(\ref{eq:asympt}) is satisfied by a class of strongly interacting QC theories described by the gauge-gravity duality.  

Working at general $z$, we first explain the origin of the $c_1$ term by using a scaling argument.   Let us imagine that 
the system is at $T>0$ in the QC fan. Since there is no phase transition in the fan,
the conductivity will receive a correction $\delta\sigma$ that is \emph{analytic} in the coupling $g$ about $g\!=\!0$, 
which generally will be linear. 
Further, by using the scaling dimension of $g$, and the fact that $\omega\gg T$ is the 
dominant energy scale, we get $\delta\sigma\sim g/\omega^{(2+z-\Delta)/z}$.    We stress that this term does \emph{not} depend on $T$. 
A more precise \bl{and general} argument can be made by first expressing the dynamical conductivity as  
$\sigma(\omega)\!=\!\tfrac{1}{\mathrm i\omega} \langle J_x J_x \mathrm{e}^{-\mathrm{i}g\int_x\! \mo}\rangle_T \mathcal Z_{0,T}/\mathcal Z_{g,T}$,
using Eq.\thinspace(\ref{eq:action}),  
where $\mathcal Z_{g,T}$ is the full partition function. 
The expectation value is taken using the $g\!=\!0$ action, and \bl{temperature $T\!\geq\! 0$}. 
We expand $\mathrm{e}^{-\mathrm{i}g\int_x\! \mo}$ to first order in $g$, and evaluate the resulting 3-point function $\langle J_x(\omega)J_x(-\omega)\mo(\tilde \omega\to 0)\rangle_{T} = \omega^{(\Delta-z-d)/z} \mathcal{F}(T/\omega)$, for a scaling function $\mathcal{F}$
(note that spatial momenta are set to zero).   Generically, $\mathcal{F}(0) \ne 0$ and is a property of the QCP at $T=0$.   
Hence, as $\omega\! \gg\! T$, $c_1=\mathcal{F}(0)$ and is $T$-independent. 
If there is no phase transition as \bl{we vary $T$ at fixed $g\!\neq\! 0$}, 
by adiabaticity $c_1$ must remain unchanged all the way to, \bl{and including}, $T\!=\!0$.        

In contrast to the $c_1$-term, the $c_2$ term depends on both $g$ and $T$ through the expectation value of $\mo$, 
and was previously identified at finite temperature but zero detuning $g\!=\!0$ (and $z\!=\!1$) \cite{Katz}.  
Let us recall the main idea of that derivation, focusing on the case $z=1$, and see how it generalizes to $g\!\neq\! 0$. 
The Kubo formula for the conductivity states that we need to evaluate the current-current
correlation function. Since we are interested in short times (large-frequencies)
we consider the operator product expansion (OPE) of $J_x(t,0) J_x(0,0)$ 
in the $t\!\to\!0$ limit. Crucially, by spacetime locality 
the product can be replaced by a sum of local operators evaluated at $t\!=\! 0$,
with increasing scaling dimensions. The first non-trivial term in the sum  will generally arise due to the leading relevant
operator at the QCP, $\mathcal O$, and will be $\sim t^{\Delta -2d} \mo(t\!=\!0)$. 
We can take the expectation value of the OPE at finite $g$ and $T$ since we work at short times, $t\!\ll |g|^{-\nu z},T^{-1}$. 
Fourier transforming then leads to the $c_2$ term in \req{asympt}.  
$c_2$ itself depends on neither $g$ nor $T$; it is related to a coefficient in the OPE. 
In contrast, the $z\neq 1$ case is not as simple due to the lack of a sharp notion of spacetime locality needed to constrain the OPE.
The $c_2$ term at $z\!\neq\! 1$ is allowed by scaling, and below we will confirm its existence in a class of interacting Lifshitz theories.

The perturbative expansion used to derive the $c_1$-term is different from the commonly used perturbative expansions about a free (Gaussian) theory:
  it uses the structure of the generally interacting QCP itself to determine  
the corrections at finite detuning. 
The expansion should hold  when the detuned system
has a finite correlation length,
but can fail in regions separated from the ``fan'' by 
a phase transition, where potentially new gapless modes can arise.   We will see an example of this failure later.  

We have obtained the asymptotic expansion Eq.~(\ref{eq:asympt}) near generic QCPs. 
In the context of \emph{classical} critical phenomena, 
similar expansions for short-distance spatial correlators of the order parameter have been found for thermal Wilson-Fisher fixed points 
in 3D (where $z\!=\!1$)  \cite{Fisher-Langer, Brezin74}.  The coefficients in the expansion for these 
spatial correlators have recently  
been computed for the strongly-coupled Ising critical point \cite{Caselle:2016mww}.   These classical results are most similar to Eq.~(\ref{eq:asympt}) analytically continued \cite{caron09,will-prl} to imaginary time, when $z\!=\!1$ and $T\!=\!0$.  In this limit, 
the asymptotic behavior of short-distance correlators contains both analytic and non-analytic terms  
in the thermal detuning parameter $(T-T_{\mathrm{c}})$, since $\langle \mathcal{O}\rangle\!\sim\! |g|^{\nu\Delta}$ where $g$ is interpreted as $(T-T_{\mathrm{c}})$
under the quantum-to-classical mapping. 
This highlights that Eq.~(\ref{eq:asympt}), just as in the classical case, cannot be derived via a single perturbative expansion. Our derivation
indeed illustrates the different mechanisms behind the $c_1$ and $c_2$ terms,
and is valid near QCPs at finite $g$ and $T$, as well as when $z\!\ne\!1$. 

{\bf Universal ratios:} For QCPs described by conformal field theories ($z\!=\!1$), 
the expansion described above to get the $c_1$ term is called \emph{conformal perturbation theory}, 
and is very powerful because the 3-point function $\langle J(x_1)J(x_2)\mo(x_3) \rangle_{\rm QCP}$ is fixed by conformal symmetry and operator dimensions up to a single theory-dependent constant.
(This is not the case for general $z$.)   
The conformal symmetry thus allows us to show that for all CFTs 
the ratio $c_1/c_2$ is universal and only depends on $\Delta$ and the normalization of $\mo$:
\begin{align} \label{eq:asympt-cft}
  \frac{c_1}{c_2} = \mathcal C_{\mo \mo} \frac{-\G(4-\De)\G(\De-\tfrac{3}{2})}{2^{6-4\De} \G(1+\De)\G(\tfrac{3}{2}-\De)}, \quad c_2=\mathcal C_{JJ\mo}, 
\end{align}
where we have given the answer in 2d. $\Gamma(x)$ is the gamma function, and $\mathcal C_{\mathcal{OO}}$
appears in the correlator
$\langle\mo(-p)\mo(p)\rangle_{\rm QCP}\!=\!\mathcal C_{\mo\mo}p^{2\De-3}$
expressed in frequency-momentum space. The real constant $\mathcal C_{JJ\mo}$ 
enters in the 3-point function $\langle JJ\mo\rangle_{\rm QCP}$. The detailed derivation of \req{asympt-cft} and
its generalization to $d\!\neq\! 2$ is given in App.\thinspace\ref{ap:cft}.    

In order to get insight about the generic $z$ case, we employ the holographic gauge-gravity 
duality \cite{Maldacena,Zaanen:2015oix, ALRMP} to study charge transport in a class of interacting large-$N$    
matrix field theories. Such theories are dual to gravitational theories existing  
in a $(d\!+\!2)$-dimensional curved spacetime whose isometries are in correspondence with the 
Lifshitz symmetries of the matrix field theories at general $z$. This approach is useful because techniques such as conformal perturbation theory, which are non-perturbative in interaction strength and robust against the large $N$ limit, are not known for $z\ne 1$.
Details of the computation will be presented in \cite{ALp3}; we give the result for $\langle JJ\mathcal{O}\rangle $ for general $d$ 
in App.\thinspace\ref{ap:lif}.   We follow the logic of conformal perturbation theory to demonstrate Eq.~(\ref{eq:asympt}) and predict $c_{1,2}$;  a direct computation of the high frequency asymptotics of $\sigma(\omega)$ using gauge-gravity duality confirms our prediction \cite{ALp3}.  In 2d, we find  
\begin{align} \label{eq:asympt-z} 
  \frac{c_1}{c_2}\! =\!  \frac{-\mathcal C_{\mo \mo}\,\G(2+\frac{2-\De}{z})\G(\frac{\Delta-1}{z}-\frac{1}{2})}{2^{\frac{2}{z}(2+z-2\De)} \G(1+\frac{\De}{z})\G(\tfrac{1}{2} + \frac{1-\De}{z})}, \;\; c_2=\mathcal C_{JJ\mo},
\end{align}
for $2\Delta\!\neq d+z +2n$, for integer $n$.    
Results for general $d$ can be found in App.~\ref{ap:lif}.   
We note that Eq.~(\ref{eq:asympt-z}) reduces to Eq.~(\ref{eq:asympt-cft}) when $z\!=\!1$. 
Unlike Eq.~(\ref{eq:asympt-cft}), the holographic result for $c_1/c_2$ at $z\!\neq\!1$ is unlikely generic. 
Indeed, $\langle J_x(\omega_1)J_x(\omega_2)\mathcal{O}(\omega_3)\rangle_{\rm QCP}$ is not sharply constrained by Lifshitz symmetry.   We do expect, however, that the asymptotic form of (\ref{eq:asympt}) remains the same near other $z\ne 1$ QCPs.    Indeed, above we have provided a general scaling argument for the $c_1$-term at any $z$.

{\bf Sum rules:}
We can use the high-frequency expansion \req{asympt} to derive sum rules for the conductivity. 
This was previously done for CFTs at finite temperature but zero detuning \cite{Gulotta,WS12,WS13,Katz}.      
At $g\!\neq\! 0$, one must take into consideration the new $c_1$ term
in the asymptotic expansion Eq.~(\ref{eq:asympt}), which will drastically change the result in many cases. 
For $d+z-2<\De<2$, the sum rule reads
\begin{align} \label{eq:sr}
  \int_0^\infty \mathrm{d}\omega \, \re\left[\sigma(\omega) - \sigma(\omega)\big|_{T=g=0} \right] = 0\,.
\end{align}
If $\De\!>\! 2$ or $\Delta\!<\! z+d-2$, the integral becomes infinite making Eq.~(\ref{eq:sr}) ill-defined.  
Thus, in contrast to $d\!=\! 2$, most states in $d\!=\! 3$ will not obey Eq.~(\ref{eq:sr}) since generally $z\!\geq\! 1$.
In the special case of 1d CFTs, the conditions on $\Delta$ for the validity of (\ref{eq:sr}) are trivially satisfied. 
For general $d$, $\Delta\!=\! 2$ or $z+d-2$ constitute special cases since the rhs of Eq.~(\ref{eq:sr}) 
can be finite and non-zero
(see the $O(N)$ model calculation below).   
Again, (\ref{eq:sr}) holds in the same regime as the asymptotic expansion, \ie for points in 
the $(g,T)$ phase diagram that can be reached from the QC region without crossing phase transitions.
Knowledge about the expansion is needed to ensure that $\sigma(\omega)$ decays sufficiently fast at large frequencies. The other ingredient is the analyticity of $\sigma$ in the upper half-plane of complex frequencies
(causality), which allows us to prove the sum rule by contour integration (App.~\ref{ap:sr}). 

{\bf $O(N)$ model:} We now examine the physics described above in the context of the 
interacting QCPs in the $O(N)$ model in 2d, which have $z=1$ and are CFTs.  
We focus on 2 cases: $N=\infty$ (which is solvable), 
and $N=2$ which describes an interacting superfluid-insulator QCP. 
These QCPs are described by a relativistic $\phi^4$-theory for an order parameter 
field $\phi_a$ with $N$ real components \cite{ZinnJustin:2002ru}: 
\begin{equation} \label{eq:ONmain}
  S = -\int \mathrm{d}^3x \left(\frac{1}{2}\partial_\mu \phi_a \partial^\mu \phi_a + r\phi_a \phi_a + \frac{u}{2N}(\phi_a\phi_a)^2\right).  
\end{equation}
This action is written in real time.   When $r$ is large, this model yields a gapped phase with unbroken $\mathrm{O}(N)$ symmetry;  when $r$ is small, $\mathrm{O}(N)$ is spontaneously broken and the low energy effective theory contains Goldstone bosons if $N\!>\! 1$. 
There are conserved currents $J^\mu_{ab} = \phi_a \partial^\mu \phi_b - \phi_b\partial^\mu \phi_a$, and our goal is to compute the corresponding conductivity.    When $1<d<3$, dimensional analysis suggests that this QCP has a relevant operator $\phi_a\phi_a$ with detuning parameter $g\sim r$.   This is qualitatively correct; in App.\thinspace\ref{ap:ON}, we precisely identify $\mathcal{O}$ and $g$ in terms of slightly different variables.

When $N=\infty$,  this model is exactly solvable through large-$N$ techniques \cite{ZinnJustin:2002ru}. 
The resulting QCP has $\nu\!=\!1$ and is thus distinct from the Gaussian fixed point at $u\!=\!0$.  
Let us begin by studying the disordered phase, which occupies the entire phase diagram
except the broken symmetry state at $T\!=\!0$ and $g\!<\! 0$.  
We obtain the following asymptotic expansion via \bl{an explicit computation} 
of the conductivity (App.\ts\ref{ap:ON})
\begin{equation}  \label{largeN-dis}
  \sigma(\omega) =  \frac{1}{16} + \frac{4g}{\mathrm{i}\omega} - \frac{\langle \mathcal{O}\rangle_{g,T}}{4N\omega^2} + \dotsb,
\end{equation}  
where $\langle \mathcal O\rangle_{g,T} = Nm^2$, with $m(g,T)$ being the detuning and temperature induced mass, 
given in App.\thinspace\ref{ap:ON}.  
Using the previously derived values $\sigma_\infty = \frac{1}{16}$, $\Delta=2$, $\mathcal{C}_{JJ\mathcal{O}} = \frac{1}{4N}$ 
and $\mathcal{C}_{\mathcal{OO}} = -16N$ \cite{Katz}, we find exact agreement with Eq.\thinspace(\ref{eq:asympt-cft}). 
Now, the $g$-linear term, although purely imaginary, alters the sum rule Eq.\thinspace(\ref{eq:sr}) from its $g\!=\!0$ form
because we have the special situation $\Delta\!=\!2$. 
Indeed, we find that the rhs of Eq.\thinspace(\ref{eq:sr}) becomes finite, $-2\pi g$, which is \emph{independent of temperature},
and changes sign across $g\!=\! 0$, see App.\thinspace\ref{ap:sr}.

The conductivity in the ordered phase at $N\!=\!\infty$, which occurs when $T\!=\!0$ and $g\! <\!0$, is qualitatively distinct. 
When the condensate is along the 1-direction $\langle \phi_1\rangle\neq 0$,   
the asymptotic conductivity for $J_{12}^\mu$ reads
\begin{equation} \label{largeN-ord}
  \sigma(\omega) = \frac{1}{16}  + \frac{64}{3\pi^2} \frac{|g|}{\mathrm{i}\omega}  \ln \frac{\omega}{\mathrm{i} |g|} 
  + \mathrm{O}\Big( \frac{1}{\omega} \Big).   
\end{equation}
We find disagreement with (\ref{eq:asympt}), which can be understood as follows:  
conformal perturbation theory was based around the 
convergence of the $g$-expansion of $\langle JJ \mathrm{e}^{-\mathrm i g\!\int \mathcal{O}}\rangle_{\mathrm{QCP}}$. 
When $g\!<\! 0$, this expansion can lead to IR divergences associated with the instability 
of the symmetric vacuum:  $\phi_a$ has obtained an expectation value in the true vacuum.   
At $N=\infty$, 
logarithmic corrections to $\sigma$ are a consequence of the coupling to Goldstone bosons, 
as we show in App.~\ref{ap:ON}. 
Deviations from Eq.~(\ref{eq:asympt}) hence follow from the superfluid instability of the symmetric vacuum when $g\!<\!0$.
We also note that the new logarithmic enhancement in Eq.~(\ref{largeN-ord}) makes the sum rule Eq.~(\ref{eq:sr}) ill-defined because
the integral diverges. \bl{Further, the logarithmic contribution in Eq.\thinspace(\ref{largeN-ord}) is present 
when $2\!<\!d\!<\!3$, for all temperatures at which long range order exists, with a proportionality coefficient related to the superfluid density (see App.~\ref{ap:ON}).}

\begin{figure}
  \center
  \includegraphics[scale=.53]{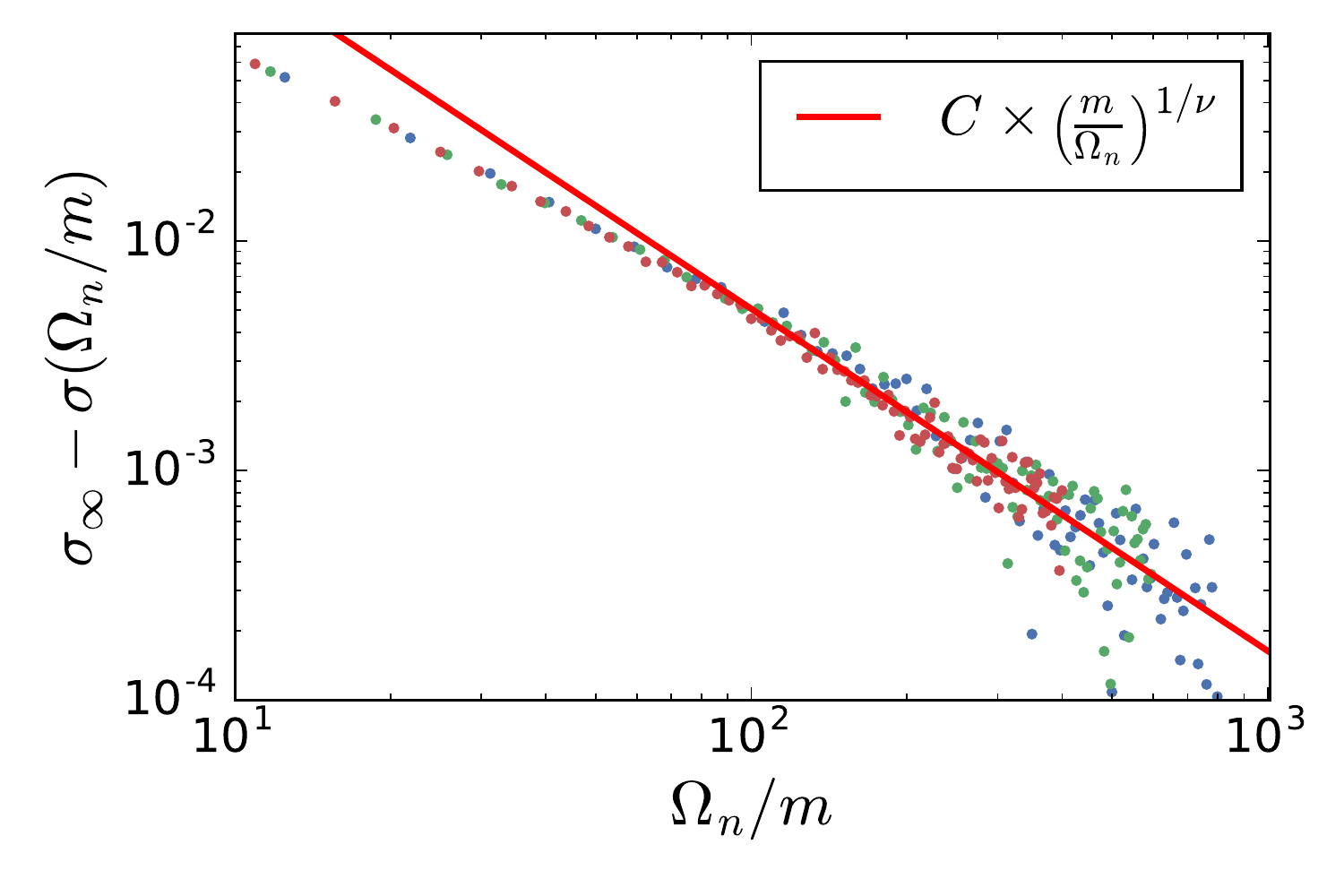} 
  \caption{Log-log plot of the asymptotic behavior of $\sigma(\mathrm i\Omega_n)$ at imaginary frequencies, in the disordered phase of the $\mathrm{O}(2)$ model, computed using QMC in the limit $T\!\to\!0$. Each set of colored dots represents a different detuning $g$. $m\!\propto\! g^\nu$ is the single particle gap.
 The line is the field theory prediction (\ref{eq:asympt}) at large $\Omega_n$, with $\nu=0.67$.}    
  \label{fig:qmc}  
  \centering   
\end{figure}

When $N\!=\!2$, the model Eq.~(\ref{eq:ONmain}) describes a strongly interacting superfluid-insulator QCP, where quasiparticle excitations have been 
destroyed by fluctuations. 
We analyze its imaginary time conductivity numerically using large-scale lattice quantum Monte Carlo (QMC) simulations. 
We work with the action Eq.~(\ref{eq:ONmain}) in Euclidean spacetime (devoid of the sign problem), 
discretized on a  $512\times 512\times 512$ cubic lattice.  
Details of the numerical methods are in App.~\ref{ap:num}. Fig.~\ref{fig:qmc} shows the universal part of the imaginary frequency conductivity
in the disordered phase at different values of the detuning, near the QCP.   We plot the conductivity
relative to its groundstate value $\sigma_\infty$ as a function of the frequency rescaled by the single-particle 
gap $m\propto g^\nu$.  In order to do so, we must subtract off a non-universal lattice correction to $\sigma$, and employ $\sigma_\infty=0.355(5)$, found with recent conformal bootstrap calculations \cite{Kos:2015mba} 
along with numerical simulations \cite{Gazit13,Gazit14,nat,Chen14,Katz}.        
The resulting data 
collapses to a single universal curve.
The large-$\omega$ field theory prediction (solid line) for the subleading term, 
which scales as $c_1\omega^{-1/\nu}$, with $\nu=0.67$, is also shown.   At $N\!=\!2$, in contrast to the $N\!=\!\infty$ case Eq.~(\ref{largeN-dis}),
the next subleading term $\propto c_2\omega^{1/\nu-3}$ comes with nearly the same exponent,
so that in practice we combine both the $c_{1,2}$ terms into a single one. 
By looking at the high frequency limit, we see that $c_1$ is negative, in agreement with our result at $N=\infty$, 
Eq.\thinspace(\ref{largeN-dis}).   The numerical data is also consistent with our predicted scaling $\sigma-\sigma_\infty \propto \omega^{-1/\nu}$, but due to the need to subtract off a large background conductivity to extract $c_1$ and $\nu$, we presently cannot perform a more quantitative analysis. 

In the superfluid phase, both the numerical and field theory analyses become complicated by the presence of the broken symmetry
and the associated strongly coupled Goldstone boson(s) (at finite $T<T_{\mathrm{c}}$, the order becomes algebraic). 
In order to analytically understand the asymptotic behavior of $\sigma(\omega)$, and the associated sum rule Eq.\thinspace(\ref{eq:sr}), one would need to 
use methods beyond what we have discussed so far.
It will be interesting to see whether the result will be similar to the $N\!=\!\infty$ case, Eq.~(\ref{largeN-ord}), 
with the associated breakdown of the sum rule. We leave this important question for the future.   

\textbf{Outlook:}  
We have determined the large-frequency optical conductivity near a QCP for a wide class of theories, \req{asympt}, 
in general dimensions. Our analysis incorporates non-thermal detuning and temperature, and thus extends
beyond the QC fan \bl{which facilitates comparison with experiments}.
This has led to a unified understanding of sum rules in the phase diagram 
near such QCPs. Interestingly, we have found that in certain superfluid phases, interacting Goldstone bosons
can qualitatively change the results. It will be of interest to analyze such effects more broadly.
Our findings can potentially be tested at QCPs  
in superconductor-insulator systems or Josephson junction arrays \cite{cha},    
and in ultra-cold atomic gases. In the latter case, the physics of the superfluid-insulator QCP has already been realized \cite{Spielman07,Chin12,Endres12}, and proposals   
for measuring the optical conductivity exist (\eg by periodic phase-modulation of the optical lattice \cite{Tokuno}).    
Finally, although this letter focused on the optical conductivity, our general techniques apply to other correlation functions.

\textbf{Acknowledgments:} 
We acknowledge useful discussions with T.~Faulkner, R.C.~Myers, S.~Sachdev, and E.S.~S\o rensen. 
A.L.\ and W.W.K.\ were funded by MURI grant W911NF-14-1-0003 from ARO.
A.L.\ was supported by the NSF under Grant DMR-1360789. 
S.G.\ received support from the Simons Investigators Program, the California Institute of Quantum Emulation, 
and the Templeton Foundation.
W.W.K.\ was funded by a postdoctoral fellowship and a Discovery Grant from NSERC, and by a Canada Research Chair.
D.P.\ was funded by the Israel Science Foundation (ISF) Grant No.\ 1839/13 
and the Joint UGC-ISF Research Grant Program under Grant No.\ 1903/14.  
This project was initiated by W.W.K.\ and S.G.\ at a Summer School in Les Houches.
Part of the work was performed at the Aspen Center for Physics, which is supported by National Science Foundation grant PHY-1066293.    

\appendix 

\begin{center} 
{\Large\bf Appendices} 
\end{center} 

\tableofcontents          

\section{Asymptotics in Conformal Field Theory}
\label{ap:cft}

In this appendix we use techniques from conformal field theory (CFT) to derive Eq.~(\ref{eq:asympt}) 
near a QCP with $z=1$. 
Conformal field theories have an enhanced symmetry group containing Lorentz transformations and scale invariance \cite{DiFrancesco:1997nk}, and describe many $z= 1$ QCPs of physical relevance.   
As we will see, this symmetry group is powerful enough to completely fix $c_1$ and $c_2$, for any CFT, in terms of a few simple numbers (operator dimensions and operator product expansion coefficients).    We denote with $D$ the spacetime dimension, $D=d+1$. In this appendix, we shall work in Euclidean (imaginary) time.        

In a CFT, a (Lorentz) scalar operator of dimension $\Delta$ has a two-point function \begin{equation}
\left\langle \mathcal{O}(x)\mathcal{O}(0)\right\rangle = \frac{C_{\mathcal{OO}}}{x^{2\Delta}}.
\end{equation}
The only free parameters are the operator normalization $C_{\mathcal{OO}}$ and scaling dimension $\Delta>(D-1)/2$ \cite{Rychkov:2016iqz}.   
For the purposes of this work, it is convenient to work in frequency-momentum space: 
\begin{equation}
  \langle \mathcal{O}(p) \mathcal{O}(-p)\rangle = \mathcal{C}_{\mathcal{OO}} p^{2\Delta-D}\,.
\end{equation}   
(In special cases like $\Delta=D/2$, logarithms can also appear.)
From the Fourier transform, one finds 
\begin{equation} \label{eq:fourier}
 \mathcal{C}_{\mathcal{OO}} = C_{\mathcal{OO}} \times  \frac{2^{D-2\De} \pi^{D/2} \G(\tfrac{D}{2} -\De)}{\G(\De)}.
\end{equation}
As $C_{\mathcal{OO}}>0$,  for many operator dimensions $\Delta$ of interest (including $\Delta\simeq 1.51$ for the   
relevant O(2)-invariant scalar operator in the $N=2$ Wilson-Fisher QCP in $D=3$), we see that $\mathcal{C}_{\mathcal{OO}}<0$.   

When detuning the system away from criticality, we want to understand how the sourced scalar field $\mathcal{O}$ modifies the conductivity.   
As explained in the main text, it will prove useful to know the momentum-space 3-point correlator \cite{Bzowski:2013sza} 
\begin{widetext}
\begin{align}
\left\langle J_x (p_1) J_x (p_2)\mathcal{O}(p_3)\right\rangle \!=\! A_{JJ\mathcal{O}}\cdot \left[I\!\left(\tfrac{D}{2}, \tfrac{D}{2}-1, \tfrac{D}{2}-1, \Delta - \tfrac{D}{2}+1\right) + \tfrac{\Delta}{2}(D-2-\Delta) I\!\left(\tfrac{D}{2} - 1, \tfrac{D}{2}-1, \tfrac{D}{2}-1, \Delta - \tfrac{D}{2} \right) \right]  \label{eq:JJOformal}
\end{align}
\end{widetext}
where $p_{1,2,3}$ are chosen to lie in the time direction:\begin{subequations}\begin{align}
p_1 &= (\Omega,0), \\
p_2 &= (-\Omega-p,0), \\
p_3 &=  (p,0),
\end{align}\end{subequations}
with $p,\Omega\geq 0$, and we have defined 
\begin{equation}
I(a,b,c,d) \equiv \int\limits_0^\infty \mathrm{d}x \; x^a p_1^b p_2^cp_3^d\, \mathrm{K}_b(p_1x) \mathrm{K}_c(p_2x)\mathrm{K}_d(p_3x) \label{eq:Idef}
\end{equation}Here $\mathrm{K}_a$ is the modified Bessel function of the second kind. 
Once again, we see that up to an overall normalization $A_{JJ\mathcal{O}}$, the form of Eq.~(\ref{eq:JJOformal}) is completely fixed by conformal invariance.   In what follows, we will focus on the 
limit $p\ll \Omega$, relevant for the computation of the high frequency conductivity. 

 The presence of the scalar operator $\mathcal{O}$ modifies the operator product expansion (OPE) associated with a conserved current.   In momentum space, the OPE of the current operator (obtained by Fourier transforming the real space 
form) contains the non-analytic term \begin{equation}
J_x(\Omega) J_x(-\Omega-p)  = \mathcal{C}_{JJ\mathcal{O}}\Omega^{D-2} \frac{\mathcal{O}(-p)}{\Omega^\Delta} + \cdots . \label{eq:CJJOmom}
\end{equation}
in the limit $p\ll \Omega$.    
The OPE coefficient $\mathcal{C}_{JJ\mathcal{O}}$ can be related to $A_{JJ\mathcal{O}}$ by contracting both sides of Eq.~(\ref{eq:CJJOmom}) with $\langle \cdots \mathcal{O}(p)\rangle$, and then taking the limit $p\rightarrow 0$.   In doing so, one finds that the leading order singular contribution in $p$ is 
\begin{multline}
\langle J_x(\Omega) J_x(-\Omega-p) \mathcal{O}(p)\rangle = \mathcal{C}_{JJ\mathcal{O}}\Omega^{D-2} \frac{\mathcal{C}_{\mathcal{OO}}p^{2\Delta-D}}{\Omega^\Delta} \\
+ \cdots + (\text{regular as $p\to 0$}) .
\end{multline}
Using the small-$x$ Taylor expansion of the Bessel function:  \begin{equation}
\mathrm{K}_b(x) = (2^{b-1} \Gamma(b) x^{-b} + \cdots) + (2^{-b-1}\Gamma(-b) x^b + \cdots),  \label{eq:bessel}
\end{equation}
we find that the $p^{2\Delta-D}$ contribution in Eq.~(\ref{eq:JJOformal}) arises from the second term in the above expansion: \begin{widetext} \begin{align}
I\left(\tfrac{D}{2}-1,\tfrac{D}{2}-1,\tfrac{D}{2}-1,\Delta-\tfrac{D}{2}\right) &=   p^{2\Delta-D}  \int\limits_0^\infty \mathrm{d}x \; x^{\Delta-1}  \Omega^{D-2} \left[ \mathrm{K}_{D/2-1}(\Omega x)\right]^2 2^{D/2-\Delta-1} \Gamma\left(\tfrac{D}{2}-\Delta\right)  + \cdots \notag \\
&= p^{2\Delta-D} \frac{\Omega^{D-2}}{\Omega^{\Delta }} 2^{D/2-\Delta-1} \Gamma\left(\tfrac{D}{2}-\Delta\right)\Psi\!\left(\Delta; \tfrac{D}{2}-1\right),   
\end{align}
with the function \begin{equation}
\Psi(a;b) \equiv \int\limits_0^\infty \mathrm{d}x \; x^{a-1} \mathrm{K}_b(x)^2 =  \frac{\sqrt{\pi}\, \Gamma(\frac{a}{2}) \Gamma(\frac{a}{2}+b) \Gamma(\frac{a}{2}-b)}{4\Gamma(\frac{1+a}{2})} .
\end{equation}
Hence, we find the relation 
\begin{equation}
  \mathcal{C}_{JJ\mathcal{O}}= -\frac{A_{JJ\mathcal{O}}}{\mathcal{C}_{\mathcal{OO}}} \Delta \left(1-\tfrac{D-\Delta}{2}\right) 2^{D/2 - 1-\Delta} 
  \Gamma\left(\tfrac{D}{2}-\Delta\right) \Psi\!\left(\Delta; \tfrac{D}{2}-1\right) .  \label{eq:ACJJO}
\end{equation}

\subsection{Conductivity}
Given the CFT data described above, we are now ready to use conformal perturbation theory to compute the asymptotic behavior of the two-point function $\langle J_x(\Omega)J_x(-\Omega)\rangle$  
when we detune away from criticality, by a finite temperature $T$, and by a coupling constant to a (relevant) scalar operator $\mathcal{O}$.   For simplicity, we assume that $\Delta \ne D/2 + n$,  where $n$ is an integer. Assuming that conformal perturbation theory is well behaved, we find \begin{equation} 
\langle J_x (\Omega) J_x(-\Omega)\rangle_{g}  =  
\frac{\mathcal Z_{g=0}}{\mathcal Z_g}\left\langle J_x (\Omega) J_x(-\Omega) \mathrm{e}^{-g \mathcal{O}(0)}\right\rangle_{\! g=0}  
= \left\langle\, J_x (\Omega) J_x(-\Omega) \left[1 -g \mathcal{O}(0) \right]\, \right\rangle_{\! g=0}^c + \cdots
\end{equation}
where $\int \mathrm d^{d+1}x\, \mathcal O(x)=\mathcal O(0)$ is the $p\to 0$ limit of the Fourier transform of $\mathcal O(x)$, and the superscript ``c'' denotes the connected correlation function, which we will omit from now on for simplicity.  
Here, we have omitted the temperature $T$ as the correction to the groundstate conductivity that we study first
will not depend on $T$ at all.
This correction is linear in $g$ and arises from the finite expectation value 
\begin{align}
 \lim_{p\to 0} \left\langle J_x (\Omega) J_x(-\Omega) \mathcal{O}(p) \right\rangle = {\rm finite}\,,
\end{align}
which is evaluated at the QCP, namely in the groundstate of the CFT.
Whenever $\Delta>D/2$, it is in fact the leading order contribution in the $p\rightarrow 0$ limit.  
Indeed, when comparing $A_{JJ\mathcal{O}}$ to $\mathcal{C}_{JJ\mathcal{O}}$, we Taylor expanded a Bessel function in Eq.~(\ref{eq:JJOformal}).  If we focused on the first term in Eq.~(\ref{eq:bessel}), we find that the $p$-dependence of the correlator drops out.   So we take the $p\rightarrow 0$ limit, and combining Eqs.~(\ref{eq:JJOformal}), (\ref{eq:Idef}) and (\ref{eq:bessel}), we find at $g=0$:
\begin{align}
&\left\langle J_x (\Omega) J_x (-\Omega)\mathcal{O}(0)\right\rangle 
= -A_{JJ\mathcal{O}}\frac{\Omega^{D-2}}{\Omega^{D-\Delta}} \cdot (D-\Delta)\left(1-\tfrac{\Delta}{2}\right) 2^{\Delta-1-D/2}\Gamma\left(\Delta-\tfrac{D}{2}\right) \Psi\left(D-\Delta;\tfrac{D}{2}-1\right).  \label{eq:AJJO}
\end{align}
This is one singular contribution to the asymptotic expansion of the current two-point function,  and another comes simply from the OPE itself, as discussed above:  
\begin{equation}
  \langle J_x(\Omega_\ell)J_x(-\Omega_\ell)\rangle_{g,T} = \left\langle \sigma_\infty \Omega_\ell^{D-2} + \Omega_\ell^{D-2} \mathcal{C}_{JJ\mathcal{O}} \frac{\mathcal{O}(0)}{\Omega_\ell^\Delta} + \cdots \right \rangle_{\! g,T}
\end{equation} 
where we have explicitly restored the temperature; $\Omega_\ell=2\pi\ell T$ is a Matsubara frequency and $\ell\geq 0$ 
an integer.
$\langle \mathcal{O}\rangle_{g,T}$ will generally depend on both $g$ and $T$.   This latter contribution to the conductivity is local, coming from the OPE (in contrast, Eq.~(\ref{eq:AJJO}) is a non-local contribution).  Putting these two equations together, we obtain 
\begin{equation}
  \sigma(\mathrm i\Omega) = \Omega^{D-3}\left[\sigma_\infty + \frac{c_1 g}{\Omega^{D-\Delta}}  + \frac{\mathcal{C}_{JJ\mathcal{O}}\langle\mathcal{O}\rangle_{g,T}}{\Omega^\Delta}+\cdots\right], 
\end{equation} 
which, upon analytic continuation to real frequencies \cite{will-prl,caron09}, gives Eq.~(\ref{eq:asympt}) from the main text, but with $z=1$.   Recall that $c_2=\mathcal{C}_{JJ\mathcal{O}}$.   Combining Eqs.~(\ref{eq:ACJJO}) and (\ref{eq:AJJO}), we see that the ratio $c_1/c_2$ is independent of $\mathcal{C}_{JJ\mathcal{O}}$, and depends only on $\mathcal{C}_{\mathcal{OO}}$, $D$ and $\Delta$:  
\begin{equation}
  \frac{c_1}{c_2} = -\frac{\mathcal{C}_{\mathcal{OO}}}{2^{D-2\Delta}} \frac{\Gamma(1+\frac{D-\Delta}{2})\Gamma(2-\frac{\Delta}{2})\Gamma(D-1-\frac{\Delta}{2})\Gamma(\frac{1+\Delta}{2})\Gamma(\Delta-\frac{D}{2})}{\Gamma(1+\frac{\Delta}{2})\Gamma(2-\frac{D-\Delta}{2}) \Gamma(\frac{D+\Delta}{2}-1)\Gamma(\frac{1+D-\Delta}{2})\Gamma(\frac{D}{2}-\Delta)}  \label{eq:ACJJOratio}
\end{equation}
In the special case $D=3$, Eq.~(\ref{eq:ACJJOratio}) simplifies to Eq.~(\ref{eq:asympt-cft}) using $\Gamma$ function identities.

\section{Results from Lifshitz Holography}
\label{ap:lif}

In this appendix, we summarize the field theoretic results obtained by studying a special class of interacting Lifshitz field theories accessible through the gauge-gravity duality.    Gauge-gravity duality maps the correlation functions of certain field theories with large $N$ matrix degrees of freedom to classical computations in various curved spacetimes in one higher 
spacetime dimension.   In the simplest case of the correspondence, a large $N$ CFT is dual to a classical gravity theory on 
anti-de Sitter (AdS) space \cite{Maldacena},  but the correspondence is now believed to be far more generic \cite{Zaanen:2015oix, ALRMP}.
In particular, there is a ``Lifshitz" geometry, with metric    
\begin{equation} \label{eq:metric}
  \mathrm{d}s^2 = \frac{\mathrm{d}r^2}{r^2} -   \frac{\mathrm{d}t^2}{r^{2z}} + \frac{\mathrm{d}\mathbf{x}^2}{r^2}
\end{equation}
where $r$ is the extra holographic dimension, and $(t,\mathbf x)$ represent the $D$ dimensional spacetime of the Lifshitz QFT. 
The isometries of Eq.~(\ref{eq:metric}) (symmetries of the metric) may be interpreted as the symmetries (translation, spatial rotation and dilatation) of a Lifshitz field theory \cite{Kachru:2008yh}.   
Classical gravity computations in such a background are believed to reproduce the correlation functions of an unknown Lifshitz field theory. 
Note that in the special case $z=1$, the metric reduces to that of AdS, and the dual QFT is conformal, and relatively well-understood \cite{Maldacena}. 

  The details of the gravity computation are beyond the scope of this letter and will be reported in \cite{ALp3}.   The computation proceeds somewhat similarly to \cite{Myers:2016wsu}. Here, we focus merely on the final results, and compare them 
to our prediction for the high frequency conductivity, \req{asympt}.   In order to fix $c_1$ and $c_2$, we must carefully study the three-point function $\langle J_x(\Omega_1) J_x(\Omega_2) \mathcal{O}(\Omega_3)\rangle$, just as we did when $z=1$.   What we find in our holographic model is that this three-point function takes a very similar form to Eq.~(\ref{eq:JJOformal}):  
\begin{align}
\langle J_x(\Omega_1)J_x(\Omega_2)\mathcal{O}(\Omega_3) \rangle &= A_{JJ\mathcal{O}}\left\lbrace z I\!\left(\frac{d+z}{2z},\frac{d+z-2}{2z},\frac{d+z-2}{2z},\frac{2\Delta-d+z}{2z}\right) \right. \notag \\
&\left. - \frac{\Delta}{2}(\Delta+2-d-z) I\!\left(\frac{d-z}{2z},\frac{d+z-2}{2z},\frac{d+z-2}{2z},\frac{2\Delta-d-z}{2z}\right)\right\rbrace.  \label{eq:JJOz}
\end{align}
with $I(a,b,c,d)$ defined in Eq.~(\ref{eq:Idef}).  We stress that this formula is only valid when the three momenta in $I(a,b,c,d)$ are entirely in the $t$ direction.   The constant $A_{JJ\mathcal{O}}$ can also be computed in terms of certain parameters of the bulk gravity description, but its value is not relevant here.

As before, we consider the limit \begin{subequations}\begin{align}
\Omega_1 &= \Omega, \\
\Omega_2 &= -\Omega-p, \\
\Omega_3 &= p,
\end{align}\end{subequations} 
with $p\ll \Omega$.  The correlator Eq.~(\ref{eq:JJOz}) has a term regular in $p$ as $p\to 0$, given by
 \begin{align}
\langle J_x(\Omega)J_x(-\Omega)\mathcal{O}(0) \rangle 
= -A_{JJ\mathcal{O}} \Omega^{\frac{\Delta-z-2}{z}} 2^{\frac{2\Delta-d-3z}{2z}} \left(1-\frac{\Delta}{2}\right)(d+z-\Delta) \Gamma\left(\frac{2\Delta-d-z}{2z}\right)\Psi\! \left(\frac{d+z-\Delta}{z} ; \frac{d+z-2}{2z}\right).
\end{align}
Similarly, we find a non-analytic contribution in $p$: \begin{align}
\langle J_x(\Omega)&J_x(-\Omega)\mathcal{O}(p) \rangle = \cdots  -  A_{JJ\mathcal{O}} \Omega^{\frac{\Delta-z-2}{z}} \left(\frac{p}{\Omega}\right)^{\frac{2\Delta-d-z}{z}}  
\frac{\Delta(\Delta+2-d-z)}{2} 2^{\frac{d-z-2\Delta}{2z}}\Gamma\left(\frac{d+z-2\Delta}{2z}\right) \Psi\! \left(\frac{\Delta}{z} ; \frac{d+z-2}{2z}\right) .
\end{align}
We attribute this non-analytic contribution to the presence of the operator $\mathcal{O}$ in the OPE of $J_x J_x$: \begin{equation}
J_x(\Omega)J_x(-\Omega) =\cdots +\frac{\mathcal{C}_{JJ\mathcal{O}}}{\Omega^{\frac{\Delta+2-d-z}{z}}}\mathcal{O}(0) + \cdots
\end{equation} where \begin{equation}
\mathcal{C}_{JJ\mathcal{O}} \equiv -\frac{A_{JJ\mathcal{O}}}{\mathcal{C}_{\mathcal{OO}}}  \frac{\Delta(\Delta+2-d-z)}{2} 2^{\frac{d-z-2\Delta}{2z}}\Gamma\left(\frac{d+z-2\Delta}{2z}\right) \Psi\!\left(\frac{\Delta}{z} ; \frac{d+z-2}{2z}\right).
\end{equation}
The extent to which such an OPE is well-behaved for general non-conformal theories is not well understood \cite{Bekaert:2011qd, Golkar:2014mwa, Goldberger:2014hca}.    Our holographic results are consistent nonetheless with this non-analytic contribution emerging from an OPE.   Following the logic of conformal perturbation theory, we hence fix the ratio 
\begin{equation}\label{eq:sec73AB} 
  \frac{c_1}{c_2} = -\frac{\mathcal{C}_{\mathcal{OO}} }{2^{(d+z-2\Delta)/z} } \frac{(2-\Delta)(d+z-\Delta)}{\Delta(2+\Delta-d-z)} \frac{\Gamma(\frac{2\Delta-d-z}{2z})\Psi(\frac{d+z-\Delta}{z}; \frac{d+z-2}{2z})}{\Gamma(\frac{d+z-2\Delta}{2z})\Psi(\frac{\Delta}{z}; \frac{d+z-2}{2z})}
\end{equation} 
\end{widetext}
In a separate calculation, we can compute the high-frequency expansion of the conductivity of the theory 
dual to Eq.~(\ref{eq:metric}), and find
that it exactly matches the result of the three-point function calculation, Eq.~(\ref{eq:sec73AB}). 
In $d=2$, this reduces to Eq.~(\ref{eq:asympt-z}). 

\section{Conductivity of the $\mathrm{O}(N)$ Model at $N=\infty$}
\label{ap:ON}

The (Euclidean time) action of the $\mathrm{O}(N)$ model is \begin{equation}
S = \frac{1}{2} \int \mathrm{d}^{d+1}x \left[ \partial_\mu \phi^a \partial^\mu \phi^a + \frac{u}{N} \left(\phi^a\phi^a  -\frac{N}{\mathfrak{g}}\right)^2 \right].  \label{eq:ONaction}
\end{equation}
with $\mu$ indices running over spacetime coordinates, and $a=1,\ldots,N$.  
There are ${N \choose 2}$ conserved currents associated with the $\mathrm{O}(N)$ global symmetry:   \begin{equation}
J^{ab}_\mu = \phi^a \partial_\mu \phi^b - \phi^b \partial_\mu \phi^a,
\end{equation}   
and it is the two point correlator of this current which we will compute.   In the limit $N= \infty$, this model becomes exactly solvable for any $d$ \cite{ZinnJustin:2002ru,CSY,ssbook}.    The solution is made manifest by performing a Hubbard-Stratonovich transformation to Eq.~(\ref{eq:ONaction}):  
\begin{align}
S = \frac{1}{2}\int\! \mathrm{d}^{d+1}x \left[\partial_\mu \phi^a \partial^\mu \phi^a + \frac{\mathrm{i}\lambda}{\sqrt{N}} \left(\phi^a \phi^a - \frac{N}{\mathfrak{g}}\right) + \frac{\lambda^2}{4u}\right]\! , \label{eq:ONaction2}
\end{align}
and taking the $u\to\infty$ limit, which imposes the constraint $\phi_a\phi_a=N/\mathfrak g$ (the theory
is then a sigma model).
For spatial dimensions $1<d<3$, this model has an interacting QCP obtained by tuning $\mathfrak g$,
and distinct from the Gaussian one at $u=0$. 
We shall study correlation functions in the vicinity of this fixed point. 

The relevant scalar operator of interest here, $\mathcal{O}$, is often crudely thought of as $\phi^2 = \phi^a\phi^a$.   However, one finds more precisely 
that at $N=\infty$ \cite{Katz}: 
\begin{equation}
  \mathcal{O} =  \mathrm{i} \sqrt N \lambda.
\end{equation}
(Our normalization of $\mathcal O$ differs by a factor of $\sqrt N$ from that in \cite{Katz}.)
We will split our discussion from henceforth into two parts, depending on whether the model is in a disordered phase where $\langle \phi^a\rangle =0$, or an ordered phase where $\langle \phi^a\rangle \neq 0$.    Let us note that in all $d$, the dimension of $\mathcal{O}$ is $\Delta=2$, and the fixed point has $z=1$.

\subsection{Disordered Phase}
In the disordered phase,
the saddle point equations imply that $\langle \phi^a \rangle = 0$, and 
\begin{equation}
  \langle \mathcal O \rangle =  N m^2 ,
\end{equation} 
where $m$ is an effective mass that depends on $\mathfrak g$ and temperature $T$, and scales as $N^0$. 
We perform the Gaussian path integral over $\phi^a$ in Eq.~(\ref{eq:ONaction2}).      Keeping only the leading order terms at $N=\infty$, one finds \cite{Podolsky:2012pv}
\begin{equation}
  S_{\mathrm{eff}} = \int \mathrm{d}^{d+1}x \left[ \mathrm{i}g\sqrt{N}\lambda +  \frac{1}{2} \lambda\, \Pi(-\partial^2)\, \lambda \right]
\end{equation}
where 
\begin{subequations}\begin{align}
g &= -\frac{1}{2\mathfrak{g}} + \frac{1}{2\mathfrak{g}_{\mathrm{c}}},  \\
\Pi(p^2) &= \frac{1}{4u} + \frac{1}{2}\int \frac{\mathrm{d}^Dq}{(2\pi)^D} \frac{1}{(p+q)^2q^2}. \label{eq:Pip2}
\end{align}\end{subequations} 
\begin{widetext} 
Hence we have identified the detuning parameter $g$ in terms of the 
deviation of $\mathfrak g$ from its critical value, $\mathfrak g_c$. 
Let us review the explicit relation between $g,m$ and $T$ in the disordered phase: 
\begin{align}
  \frac{1}{\mathfrak{g}} - \frac{1}{\mathfrak{g}_{\mathrm{c}}} &\equiv \int \frac{\mathrm{d}^d\mathbf{k}}{(2\pi)^d} T\sum_{\omega_n} \frac{1}{|\mathbf{k}|^2+\omega_n^2+m^2}  - \int \frac{\mathrm{d}^{d+1}p}{(2\pi)^{d+1}}\frac{1}{p^2} \notag \\
  &= \int \frac{\mathrm{d}^d\mathbf{k}}{(2\pi)^d} \left[\frac{\coth(\epsilon(|\mathbf{k}|)/2T)}{2\epsilon(|\mathbf{k}|)} - 
    \frac{1}{2|\mathbf k|}\right] = \int \frac{\mathrm{d}^d\mathbf{k}}{(2\pi)^d} \left[\frac{n_{\textsc{b}}(\epsilon(|\mathbf{k}|))}{\epsilon(|\mathbf{k}|)} + \frac{1}{2\sqrt{\mathbf{k}^2+m^2}}- \frac{1}{2|\mathbf{k}|}\right] \notag \\
  &= \int \frac{\mathrm{d}^d\mathbf{k}}{(2\pi)^d} \frac{n_{\textsc{b}}(\epsilon(|\mathbf{k}|))}{\epsilon(|\mathbf{k}|)}  + \frac{m^{d-1}}{(4\pi)^{(d+1)/2}}\Gamma\left(\frac{1-d}{2}\right) \notag \\
  &= \frac{1}{2^{d-1}\pi^{d/2}\Gamma(\frac{d}{2})} \int_m^\infty \mathrm{d}\epsilon \left(\epsilon^2-m^2\right)^{\frac{d-1}{2}} \frac{n_{\textsc{b}}(\epsilon)}{\epsilon}  + \frac{m^{d-1}}{(4\pi)^{(d+1)/2}}\Gamma\left(\frac{1-d}{2}\right).  \label{eq:1g1gc}
\end{align}
where, here and below, we have defined the single particle dispersion relation
\begin{equation}
  \epsilon(|\mathbf k|)\equiv \sqrt{|\mathbf k|^2+m^2},
\end{equation}
as well as the Bose-Einstein distribution, $n_{\textsc{b}}(\epsilon)=1/(e^{\epsilon /T}-1)$.
$\omega_n=2\pi n T$ is a bosonic Matsubara frequency with $n$ being an integer. 
In $d=2$, one finds the closed form solution \cite{ssbook}
\begin{align}
  m(g,T) = 2T \sinh^{-1}\left( \tfrac{1}{2}\mathrm{e}^{4\pi g/T} \right).   
\end{align}
We must also compute the normalization of the two point function $\mathcal{C}_{\mathcal{OO}}$, in order to compare our direct computation of $c_{1,2}$ with Eq.~(\ref{eq:ACJJOratio}).   From Eq.~(\ref{eq:Pip2}),  it is straightforward to see 
\begin{equation}
  -\frac{N}{\mathcal{C}_{\mathcal{OO}}} \equiv p^{4-D} \int\frac{\mathrm{d}^Dq}{(2\pi)^D} \frac{1}{2q^2(q-p)^2} 
  = \frac{\Gamma(\frac{d-1}{2})\Gamma(\frac{3-d}{2})}{2^{2d}\pi^{d/2}\Gamma(\frac{d}{2})}. 
  \label{eq:CJJOON}
\end{equation}
Standard Feynman tricks may be used to compute this integral. 

The conductivity follows directly from coupling the action Eq.~(\ref{eq:ONaction}) to a gauge field $A^{12}_\mu$ and computing $\frac{\delta^2 S}{\delta (A_x^{12} )^2}$, as described in \cite{damle1997}: 
\begin{equation}
  \sigma(\mathrm i\Omega_\ell) = -\frac{T}{\Omega_\ell}\sum_{\omega_n} \int \frac{\mathrm{d}^d\mathbf{k}}{(2\pi)^d} \left[ \frac{4}{d}|\mathbf{k}|^2 
G_{\phi_2\phi_2}(\Omega_\ell - \omega_n,\mathbf{k})G_{\phi_1\phi_1}(\omega_n,-\mathbf{k}) - G_{\phi_1\phi_1}(\omega_n,\mathbf{k})-G_{\phi_2\phi_2}(\omega_n,\mathbf{k})\right] \label{eq:s12}
\end{equation} 
where $\Omega_\ell = 2\pi \ell T$ is a Matsubara frequency ($\ell\geq 0$ is an integer),
and the $\phi$ Green's function in the disordered phase is 
\begin{equation} \label{eq:G_phi}
  G_{\phi_1\phi_1}(\omega_n,\mathbf{k})=G_{\phi_2\phi_2}(\omega_n,\mathbf{k}) = \frac{1}{\omega_n^2+|\mathbf{k}|^2+m^2}. 
\end{equation}
As it stands, the integral Eq.~(\ref{eq:s12}) is divergent.   In the disordered phase, it may be regulated by multiplying the last two terms by $\partial k_x/\partial k_x$, and integrating by parts on $k_x$.   The sum over Matsubara frequencies may subsequently be performed explicitly, as in \cite{Katz}, along with the angular integral over $k$: 
\begin{equation}
  \sigma(\Omega_\ell ) = -\frac{1}{2^{d-3}\pi^{d/2}d\Gamma(\frac{d}{2})\Omega_\ell}\int_m^\infty \mathrm{d}\epsilon \left(\epsilon^2-m^2\right)^{d/2}\left[\frac{2n_{\textsc{b}}(\epsilon)}{\Omega_\ell^2+4\epsilon^2} - \frac{\Omega_\ell^2 }{4\epsilon^2(4\epsilon^2+\Omega_\ell^2)} - \frac{n_{\textsc{b}}(\epsilon)^2}{2T\epsilon} - \frac{(1+\epsilon/T)n_{\textsc{b}}(\epsilon)}{2\epsilon^2}\right].   \label{eq:B4}
\end{equation}
To analyze the asymptotics, we proceed in two steps.   We first begin with the second term in the above integral:
\begin{align}
\int\limits_m^\infty \mathrm{d}\epsilon &\left(\epsilon^2-m^2\right)^{d/2} \frac{\Omega_\ell}{4\epsilon^2\left(4\epsilon^2+\Omega_\ell^2\right)} = \frac{m^{d-1}}{4d\sqrt{\pi}\Omega_\ell^3} \Gamma\left(1+\frac{d}{2}\right)\Gamma\left(\frac{1-d}{2}\right)\left\lbrace d\Omega_\ell^2 \left(1+\mathrm{O}\left(\frac{m^2}{\Omega_\ell^2}\right)\right) \right.\notag \\
&\left.-2^{1-d}\frac{\sqrt{\pi}\Gamma(\frac{1+d}{2})}{\Gamma(\frac{d}{2})} \left(\frac{\Omega_\ell}{m}\right)^{d}\Omega_\ell m \left(1+2d\frac{m^2}{\Omega_\ell^2} + \mathrm{O}\left(\frac{m^4}{\Omega_\ell^4}\right)\right) \right\rbrace  \label{eq:B5}
\end{align} 
This expression contains all contributions to this integral, up to subleading polynomial contributions 
in $m/\Omega_\ell$, as denoted explicitly.   The first line of this equation contains a contribution to the conductivity at $\mathrm{O}(\Omega_\ell^{-1})$, which will be a part of the $c_1$ term;  the second line contains the $\sigma_\infty$ and $c_2$ terms respectively.
There is a second contribution to the conductivity of importance in our asymptotic expansion, which will arise from the last two terms in Eq.~(\ref{eq:B4}), and also contributes to the $c_1$ term:  
\begin{align}
  \int\limits_m^\infty \mathrm{d}\epsilon &\left(\epsilon^2-m^2\right)^{d/2} \left[\frac{n_{\textsc{b}}(\epsilon)^2}{2T\epsilon} + \frac{(1+\epsilon/T)n_{\textsc{b}}(\epsilon)}{2\epsilon^2}\right] = -\frac{1}{2} \int\limits_m^\infty \mathrm{d}\epsilon \left(\epsilon^2-m^2\right)^{d/2} \frac{\partial}{\partial \epsilon} \frac{n_{\textsc{b}}(\epsilon)}{\epsilon} \notag \\
  &= \frac{d}{2} \int\limits_m^\infty \mathrm{d}\epsilon \left(\epsilon^2-m^2\right)^{\tfrac{d}{2}-1 }
  n_{\textsc{b}}(\epsilon) .  \label{eq:B6}  
\end{align}
Combining Eqs.~(\ref{eq:1g1gc}), (\ref{eq:B5}) and (\ref{eq:B6}), and using $\Gamma$-function identities, we obtain 
\begin{align}
  \sigma(\mathrm i \Omega_\ell ) &= \frac{\pi^{1-\frac{d}{2}}}{2^{2d}\Gamma(1+\frac{d}{2})\sin(\frac{\pi}{2}(d-1))} 
\left[1+2d \frac{m^2}{\Omega_\ell^2 } \right]+ \frac{2}{\Omega_\ell}\left[ \frac{1}{\mathfrak{g}}-\frac{1}{\mathfrak{g}_{\mathrm{c}}}\right] +\cdots \notag \\
  &= \frac{\pi^{1-\frac{d}{2}}\, \Omega_\ell^{d-2}}{2^{2d}\Gamma(1+\frac{d}{2})\sin(\frac{\pi}{2}(d-1))} 
\left[1+2d \frac{\langle \mathcal{O}\rangle_{g,T} }{N\Omega_\ell^2}\right] 
  - \frac{4g}{ \Omega_\ell}+\cdots.  \label{eq:sigmaDON}
\end{align} 
We emphasize that the last term, $-4g/\Omega_\ell$, does not depend on temperature,
which is a non-trivial consequence of the self-consistency equation for the mass $m(g,T)$, (\ref{eq:1g1gc}).  

Let us now compare Eq.~(\ref{eq:sigmaDON}) to the CFT formalism developed previously.   From Eq.~(\ref{eq:sigmaDON}) we have \begin{equation}
\frac{c_1}{c_2}  = -\frac{2^{2d}\Gamma(\frac{d}{2})\sin(\frac{\pi}{2}(d-1))}{\pi^{1-\frac{d}{2}}}.  \label{eq:ratio1}
\end{equation}   
On the other hand from Eq.~(\ref{eq:ACJJOratio}), using $\Delta=2$, along with Eq.~(\ref{eq:CJJOON}), 
we predict from general CFT methods that  \begin{align}
\frac{c_1}{c_2} &= \frac{2^{d+3}\pi^{d/2}\Gamma(\frac{d}{2})}{\Gamma(\frac{d-1}{2})\Gamma(\frac{3-d}{2})}  \frac{ \Gamma(\frac{d+1}{2})\Gamma(d-1)\frac{\sqrt{\pi}}{2}\Gamma(\frac{3-d}{2})}{\Gamma(\frac{5-d}{2})\Gamma(\frac{d+1}{2})\Gamma(\frac{d}{2})\Gamma(\frac{d-3}{2})} = \frac{2^{d+2}\pi^{\frac{d-1}{2}} \Gamma(\frac{d}{2})\Gamma(d-1)}{\Gamma(\frac{d-1}{2}) \Gamma(\frac{d}{2})}\sin\frac{\pi (d-3)}{2}  \label{eq:ratio2}
\end{align}
Applying a few more $\Gamma$ function identities, one finds that Eqs.~(\ref{eq:ratio1}) and (\ref{eq:ratio2}) are exactly the same.   This serves as a highly non-trivial check of our CFT formalism.    In $d=2$, it was computed in \cite{Katz} that \begin{equation}
\mathcal{C}_{JJ\mathcal{O}} = \frac{1}{4N} + \mathrm{O}\left(N^{-2}\right),
\end{equation}
and so in fact from Eq.~(\ref{eq:sigmaDON}) (in $d=2$) we see that $c_{1,2}$ agree precisely with CFT predictions.

In $d=2$, it is rather non-trivial that Eq.~(\ref{eq:sigmaDON}) agrees with Eq.~(\ref{eq:asympt}).   The reason is that, a priori, one might have expected two contributions to the $(m/\Omega)^2$ contribution to Eq.~(\ref{eq:asympt}):  one from $\langle \mathcal{O}\rangle /\Omega^2$, and one from $(g/\Omega)^2$.   Evidently, the latter contribution vanishes, implying that there is no contribution to $\sigma(\omega)$ from conformal perturbation theory at second order.     It would be interesting if there is a deep reason why this must occur in the large $N$ limit.

\subsection{Ordered Phase}
Let us briefly discuss the nature of the conductivity in the ordered phase, which requires $g<0$. For simplicity, 
we work at $T=0$, and will comment on the extension to $T>0$ briefly at the end of this subsection.  We also assume that the symmetry breaking is oriented along the $a=1$ direction, 
\begin{subequations}\begin{align}
  \langle \phi_1\rangle &= \sqrt N \varrho_0 \\
  \varrho_0 &= (-2g)^{1/2} = \left( \frac{1}{\mathfrak g}- \frac{1}{\mathfrak g_{\mathrm{c}}} \right)^{\! 1/2} , \quad \mathfrak g < \mathfrak g_{\mathrm{c}},  \label{eq:gcsf}
\end{align}\end{subequations}
so that we can write
\begin{align}
  \phi_a(x) =\left(\sqrt N \varrho_0 + \varrho(x),\, \phi_{a>1}(x) \right)
\end{align}
Our key result will be the emergence of logarithmic corrections to $\sigma(\omega)$ (corresponding to 
a current that mixes with the $a=1$ direction), which goes beyond the result given in Eq.~(\ref{eq:asympt}). 
Extending the derivation in \cite{Podolsky:2012pv} to general $d$, we find that the $\varrho$ Green's function is
\begin{equation} 
\frac{1}{G_{\varrho\varrho}(k)}  = k^2 + \frac{2\varrho^2_0 }{\Pi(k)} =  k^2 + M^{d-1} k^{3-d}, \qquad 
M^{d-1} = \frac{|\mathcal{C}_{\mathcal{OO}}|\varrho_0^2}{N}\,,   \label{eq:Grho}
\end{equation}
where we have introduced a mass scale $M$ that is associated with amplitude fluctuations of $\phi_a$ (along $a=1$). 
For instance in $d=2$, $G_{\varrho\varrho}^{-1}=k(k+M)$.
Using Eq.~(\ref{eq:s12}) at $T=0$ to compute the conductivity associated with $J_\mu^{12}$, we find 
at large $\Omega$:
\begin{equation}
  \sigma = -\frac{1}{\Omega} \int \frac{\mathrm{d}^{d+1}k}{(2\pi)^{d+1}} \, 4k_x^2 
G_{\varrho\varrho}(-k_0, -\mathbf{k}) G_{\phi_2\phi_2}(k_0-\Omega, \mathbf{k})  
+ \mathrm{O}\left(\frac{1}{\Omega}\right). \label{eq:C22} 
\end{equation}
$G_{\phi_2 \phi_2}(k) = 1/k^2 $ is simply the free massless Goldstone propagator.  
This integral is divergent, as was Eq.~(\ref{eq:s12}) in the disordered phase.    The simple method that we used to regulate Eq.~(\ref{eq:s12}) in the disordered phase fails in the ordered phase, and so we resort to a hard momentum cutoff $\Lambda$.    This leads to UV divergences in $\Lambda$ which must be subtracted away;  although such a regulator cannot unambiguously fix $\sigma$ 
at $\mathrm{O}(\Omega^{-1})$, we will be able to determine exactly the leading logarithmic correction to $\sigma$.    
The UV divergent part of Eq.~(\ref{eq:C22}) can be identified using asymptotic techniques: \begin{align}
\sigma  &= -\frac{\pi^{\frac{d-1}{2}}}{(2\pi)^d\Gamma(\frac{1+d}{2})\Omega} \int\limits_\Omega^\Lambda \mathrm{d}k \frac{4k^d}{d+1} \frac{1}{k^2 + M^{d-1}k^{3-d}} + \mathrm{O}\left(\Lambda^0\right) \notag \\
&=  -\frac{4\pi^{\frac{d-1}{2}}}{(2\pi)^d(d^2-1)\Gamma(\frac{1+d}{2})\Omega} \left[\Lambda^{d-1} - (d-1) M^{d-1}\log\frac{\Lambda}{\Omega}\right] + \mathrm{O}\left(\Lambda^0\right).   \label{eq:LambdasigmaSF}
\end{align}
Upon regularization, which involves subtracting a function $\Lambda^{d-1}\mathcal{F}(\Lambda/M)$ from $\langle J^{12}_x J^{12}_x\rangle$ (the precise form of $\mathcal{F}$ is not necessary for the present computation), we obtain a logarithmic correction to $\sigma$:
\begin{equation}
\sigma(\mathrm i\Omega) = \cdots - \frac{4\pi^{\frac{d-1}{2}}}{(2\pi)^d(d+1)\Gamma(\frac{1+d}{2})} \frac{M^{d-1}}{\Omega}\log \frac{\Omega}{M} + \cdots.  \label{eq:sigmaSF}
\end{equation}

Let us evaluate these coefficients explicitly in the special case $d=2$.  First, 
we can relate $M$ to $g$ explicitly \cite{Podolsky:2012pv}: \begin{equation}
M = 32|g|\,.
\end{equation}
Using $G_{\varrho\varrho} = 1/[k(k+M)]$: 
\begin{equation}
\sigma = -\frac{1}{\Omega}\int \frac{\mathrm{d}^3k}{(2\pi)^3} \left[\frac{4k_x^2}{k(k+M)(k_x^2+k_y^2+(k_0-\Omega)^2)} - \frac{1}{k^2} - \frac{1}{k(k+M)}\right].
\end{equation} After doing angular integrals (integrate over $\cos\theta$ variable for first term) we find \begin{equation}
\sigma = \frac{1}{\Omega} \int \frac{\mathrm{d}k }{2\pi^2}  \left[ \frac{(k^2-\Omega^2)^2\log \frac{(k+\Omega)^2}{(k-\Omega)^2} - 2k\Omega (k^2+\Omega^2)}{8\Omega^3(k+M)} + 1 + \frac{k}{k+M} \right]
\end{equation}

Taylor expanding this integrand, it is straightforward to identify a linear and logarithmically divergent contribution in $\Lambda$.   After regularization, one finds\begin{equation}
\sigma= \cdots - \frac{2M}{3\pi^2\Omega}\log\frac{\Omega}{M}+ \cdots,
\end{equation} 
in agreement with Eq.~(\ref{eq:sigmaSF}), and leading to Eq.~(\ref{largeN-ord}) in the main text.

For a current such as $J^\mu_{23}$ -- which does not mix with the direction of broken symmetry -- both propagators in Eq.~(\ref{eq:s12}) are $1/k^2$.   Hence, we find that the conductivity at all frequencies is \begin{equation}
\sigma^{(23)} = \frac{1}{16}.
\end{equation}
This again disagrees with Eq.~(\ref{eq:asympt}), and is a consequence of the breakdown of the conformal perturbative expansion
in the symmetry broken phase, as discussed in the main text.

Finally, let us briefly mention the generalization of (\ref{eq:sigmaSF}) to finite temperature $T$. 
We will consider $2<d<3$. Then it is well-known that superfluidity exists at finite temperature
$T<T_{\mathrm{c}}$, and within the superfluid phase $\varrho_0^2 \propto M^{d-1}$  
will acquire temperature dependence, i.e.\ $\varrho_0(g,T)$. The explicit form could be computed from (\ref{eq:1g1gc}) and (\ref{eq:gcsf}).  
In order to compute $\sigma(\rm i\Omega)$, we must first compute the polarization function $\Pi(\mathbf{p},\Omega_n)$. 
Generalizing (\ref{eq:Pip2}) to $T>0$, we find \begin{align}
\Pi(\mathbf{p},\Omega_n) &= \frac{1}{2}T\sum_{\Omega_\ell} \int \frac{\mathrm{d}^d \mathbf{q}}{(2\pi)^d} G_{\phi_2\phi_2}(\mathbf{p}-\mathbf{q},\Omega_\ell - \Omega_n) G_{\phi_2\phi_2}(\mathbf{q},\Omega_\ell) \notag \\
&= \frac{1}{2} \int \frac{\mathrm{d}^d \mathbf{q}}{(2\pi)^d} \frac{\left(1+2n_{\textsc{b}}(|\mathbf{q}|)\right)\left(\Omega_n^2 + |\mathbf{p}-\mathbf{q}|^2 - |\mathbf{q}|^2 \right)}{|\mathbf{q}| \left(4|\mathbf{q}|^2 \Omega_n^2 + \left(\Omega_n^2 + |\mathbf{p}-\mathbf{q}|^2 - |\mathbf{q}|^2 \right)^2\right)}
\end{align}
The integral over $\mathbf{q}$ is convergent, but we will not find it necessary to compute it analytically.   As we expect, upon sending $T\rightarrow 0$ (so that $n_{\textsc{b}}\rightarrow 0$), we may analytically recover the $T=0$ result (\ref{eq:CJJOON}).    At finite $T$, when the momenta $\mathbf{p}$ and $\Omega_n$ are large, we find \begin{equation}
\Pi(\mathbf{p},\Omega_n ) = \frac{|\mathcal{C}_{\mathcal{OO}}|}{N} \left(\Omega_n^2 + \mathbf{p}^2\right)^{\frac{d-3}{2}} +  \frac{2\pi^{\frac{d}{2}}\Gamma(d-1)}{\Gamma(\frac{d}{2})}\frac{T^{d-1}}{\Omega_n^2 + \mathbf{p}^2}  + \mathrm{O}\left(T^{d+1}\right), \;\;\; (\mathbf{p}^2+\Omega_n^2 \gg T^2).
\end{equation}
Hence, we may approximate $\Pi(p)$ with its $T=0$ form, so long as the argument $p$ is large;  corrections will arise at $\mathrm{O}((T/p)^{d-1})$.   From (\ref{eq:Grho}), we also conclude that $G_{\varrho\varrho}(\mathbf{p},\Omega_n)$ is well approximated by its $T=0$ form so long as $\mathbf{p}^2+\Omega_n^2\gg T^2$: \begin{equation}
G_{\varrho\varrho}(\mathbf{p},\Omega_n)  =  k^2 \left[1 + \left(\frac{M}{k}\right)^{d-1} - \frac{|\mathcal{C}_{\mathcal{OO}}|}{N} \frac{2\pi^{\frac{d}{2}}\Gamma(d-1)}{\Gamma(\frac{d}{2})} \left(\frac{MT}{k^2}\right)^{d-1} + \cdots\right], \qquad k^2 \equiv \mathbf{p}^2 + \Omega_n^2,  
\end{equation}
where we recall that $M$ now depends on both $g$ and $T$.
An asymptotic analysis reveals that the $\Lambda$-dependence of (\ref{eq:LambdasigmaSF}), appropriately generalized to $T>0$, is unchanged.   Hence, after regulation, we conclude that the coefficient of the logarithmic divergence in (\ref{eq:sigmaSF}) is unchanged.   The vanishing of the logarithm at the critical temperature $T=T_{\mathrm{c}}$ is due to the vanishing of the superfluid density, i.e.\ $\varrho_0,M\to 0$ as $T\to T_{\mathrm{c}}$.

\end{widetext}

\section{Sum Rules at $\Delta=2$ or $d+z-2$} \label{ap:sr}

\begin{figure}
  \centering%
  \includegraphics[scale=.2]{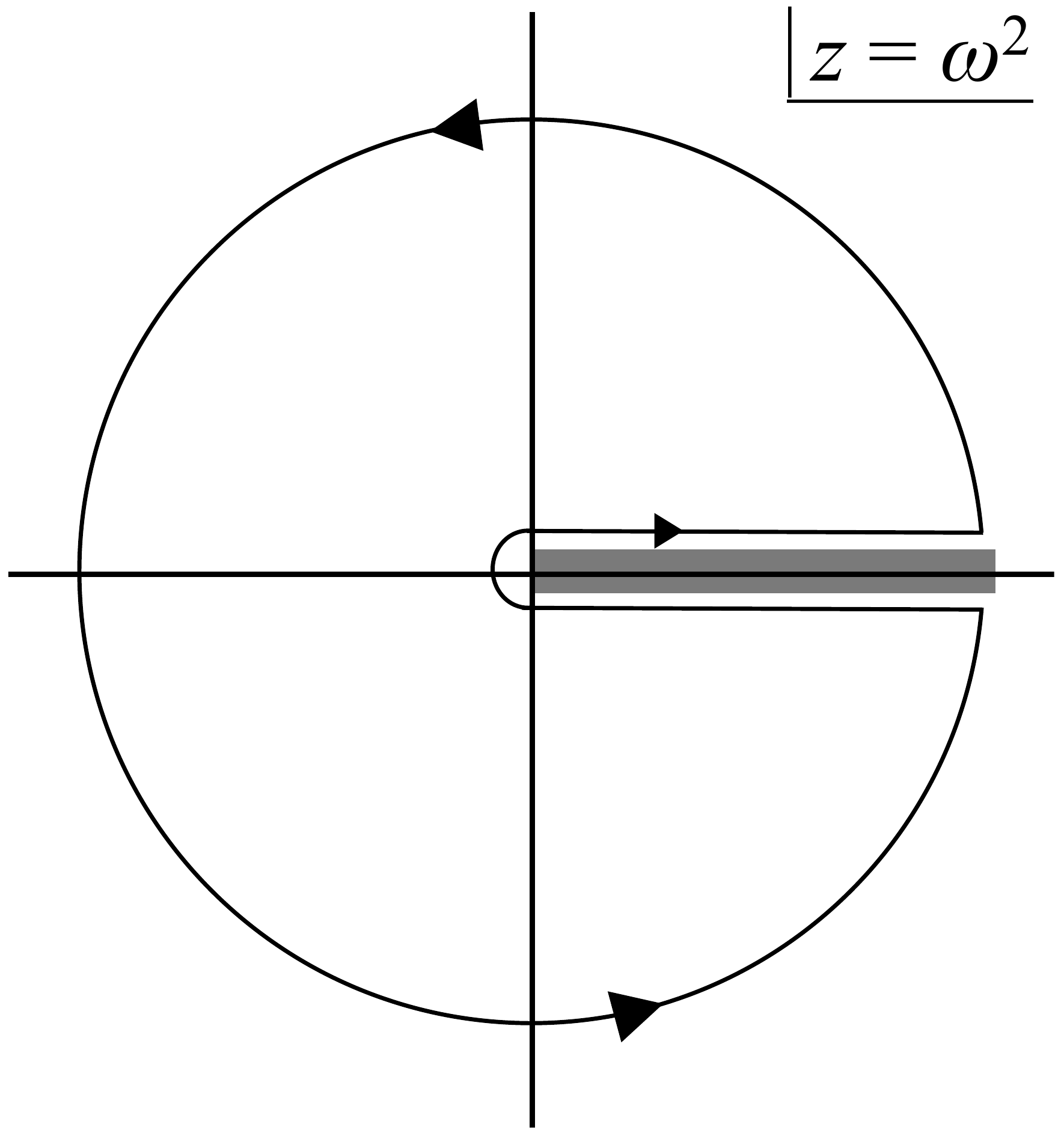}       
  \caption{\label{fig:contour} Integration contour in the complex $z=\w^2$ plane; the radius of the circle is taken
to be arbitrarily large.  
$F(z)$, defined in \req{F}, is analytic
    everywhere except on the non-negative real axis, as indicated by the thick gray line.}   
\end{figure}
We derive the sum rule for the case $\Delta=2$ or $\Delta=z+d-2$ in general dimensions
based on the asymptotic expansion and the causal properties of the current correlation function. 
In particular, when one of the conditions above is met the asymptotic conductivity will contain a term $\propto 1/(\mathrm i\omega)$. 
To be concrete, let
us consider the case $\Delta=2$ (which applies to the O$(N)$ model at $N=\infty$) so that
\begin{align} \label{eq:Delta2}
  \sigma(\omega) = (\mathrm i\omega)^{(d-2)/z} \sigma_\infty  +  \frac{c_1 g}{\mathrm i\omega} +\dotsb,
\end{align}
where the dots denote subleading terms at large $\omega$. 
Define $C_R(\omega)$ to be the retarded 2-point function of $J_x$, so that 
\begin{align}  \label{eq:C_R}
  \sigma(\omega) = \frac{C_R(\omega)}{\mathrm i\omega - 0^+}.
\end{align}
We now then introduce $\delta C_R(\omega)$ by subtracting from $C_R$ its large-$\omega$ divergence:
\begin{align}
  \delta C_R(\omega) = C_R(\omega) - (\mathrm i \omega)^{(d+z-2)/z}\sigma_\infty \,.
\end{align}
The second term is proportional to the groundstate conductivity at the QCP. 
At finite detuning $g\!\neq\! 0$, $\delta C_R(\omega)$ contains a finite term
at large $\omega$, $\delta C^\infty$; from \req{Delta2}, we see that 
\begin{align}
  \delta C^\infty= \lim_{\omega\to \infty} \delta C_R(\omega) = c_1 g\,.
\end{align}
The new function
\begin{align} \label{eq:F}
  F(\omega^2) = \delta C_R(\omega) - \delta C^\infty 
\end{align} 
is thus seen to vanish as $|\omega|\to\infty$.
We emphasize the $\omega^2$ used in the argument of $F$. 
(Given a complex number $z=|z|\mathrm{e}^{\mathrm i\theta}$, $0\leq \theta<2\pi$, we employ the convention $F(z)=\delta C_R(|z|^{1/2}\mathrm{e}^{\mathrm{i}\theta/2})-\delta C^\infty$,
with $|z|^{1/2}>0$.)  
$C_R(\omega)$ and $\delta C_R(\omega)$ are analytic in the upper half-plane $\im \omega>0$, meaning that $F(z)$ is analytic 
in  the complex $z$-plane except on the half-line $[0,\infty)$. To derive the sum rule, we consider the contour
integral 
\begin{align}
  \oint \frac{\mathrm{d}z}{2\pi \mathrm i} \frac{F(z)}{z+\alpha^2} = F(-\alpha^2)
\end{align}
where the contour is shown in Fig.~\ref{fig:contour}. Changing variables to $\omega$, where $z=\omega^2$, and using $F(|z|\to\infty)=0$,
we find
\begin{align}
  \int_0^\infty \mathrm d\omega\,\frac{2\omega}{\pi}\,\frac{\im[\delta C_R(\omega)-\delta C^\infty]}{\omega^2+\alpha^2} = F(-\alpha^2)
\end{align} 
Taking the $\alpha\to 0$ limit, we have $F(0)=\delta C_R(0)-\delta C^\infty=C_R(0)-\delta C^\infty$, which leads to
\begin{align} \label{eq:sr2}
  \int_0^\infty \mathrm d\omega\, \frac{\im \delta C_R(\omega)}{\omega} = \frac{\pi}{2} [C_R(0)-\delta C^\infty] 
\end{align}
where we used the fact that $\delta C^\infty$ is real. 
If $C_R(0)$ doesn't vanish, it will contribute a delta function $\delta(\omega)$ to $\re\sigma$.
Moving $C_R(0)$ to the l.h.s., we thus obtain the sum rule:
\begin{align} \label{eq:De2-sr}
 \int_0^\infty \mathrm d \omega \left[\re\sigma(\omega) - \sigma(\omega)\big|_{g=T=0} \right] &= -\frac{\pi}{2} \delta C^\infty\nn
  &= -\frac{\pi}{2} c_1 g\,,
\end{align} 
which is our main result. In the second equality, we have specialized the general result to the $\Delta=2$ case. 

\subsection{O$(N)$ model at $N=\infty$}
We now apply the sum rule \req{De2-sr} to the O($N$) model at $N\!=\!\infty$ in $1<d<3$:
\begin{align} \label{eq:ON-sr}
  \int_0^\infty \mathrm d \omega \re\left[\sigma(\omega) - \sigma(\omega)\big|_{g=T=0} \right] = -2\pi g\,, 
\end{align}
which holds everywhere in the phase diagram except in the ordered phase, $g<0$ at $T\leq T_{\mathrm{c}}$. 
We have used the fact that for all dimensions $1<d<3$: $\Delta\!=\! 2$, and $c_1=4$, \req{sigmaDON}.  
The special case of \req{ON-sr} at $g=0$ and $d=2$ was first derived in \cite{WS12}.   

Let us briefly comment on $C_R(0)$. It will vanish at $T=0$ when $g>0$ because
the DC conductivity vanishes. However, at $T>0$ in the O$(N)$ model 
at $N=\infty$, $\re \sigma(\omega)$ will receive a $\delta(\omega)$ contribution due to
thermally activated charge carriers, with weight proportional to $C_R(0)\neq 0$. 
This delta function is a peculiarity of the $N=\infty$ limit, where quasiparticles exist, and is not
expected at finite $N$ or more generally in interacting QCPs. 
 
\section{Monte Carlo simulations} \label{ap:num}  

\subsection{Model and observables}   
	For numerical simulations, we study  a complex scalar  $|\psi|^4$ field theory regularized on a cubic lattice. Explicitly, we consider the classical partition function $\mathcal{Z}=\int \mathcal{D}\psi \mathcal{D}\psi^*\mathrm{e}^{-\mathcal{S}[\psi,\psi^*]}$ with lattice action,
	\begin{equation}
	\mathcal{S}=\sum_{\langle i,j\rangle} \psi_i\psi_j^* +c.c +2r\sum_i|\psi_i|^2+ 4u \sum_i |\psi_i|^4 .
	\label{eq:QMC_model}
	\end{equation}
	Here, $\psi_i$ is a complex scalar field residing on the sites of a cubic lattice, which corresponds to a $D=2+1$ dimensional discretized Euclidean space-time. We study lattices with space-time volume $V=\beta\times L\times L$. Throughout, we set the inverse temperature $\beta=L$.  
The lattice model has a global $\mathrm{U}(1)$ symmetry and hence it is expected to be described at long distances
by the $\phi^4$ theory in Eq.~\eqref{eq:ONmain} with $N=2$. 
	
	At a critical coupling $u=u_c$ the system undergoes a phase transition between a disordered, $u>u_c $, and a broken symmetry phase, $u<u_c$. We define the dimensionless detuning parameter as $\delta u=\frac{u-u_c}{u_c}$.  Here, we focus only on the disordered phase, \ie $\delta u>0$.  
	Our main observable is the dynamical conductivity $\sigma(\mathrm i\Omega_n$), evaluated at Matsubara frequency $\Omega_n=2\pi n/\beta$ with $n\in \mathbb{Z}$. To define the conductivity, we introduce an external $\mathrm{U}(1)$ gauge field $A_{i,i+\eta}$ through a Peierls substitution $\psi_i\psi^*_{i+\eta}\to\psi_i\psi_{i+\eta}^*e^{\mathrm i A_{i,i+\eta}}$. The bond current is then $J_{i,i+\eta}=\frac{\delta \mathcal{S}}{\delta A_{i,i+\eta}}$ and the conductivity is defined as,
	\begin{subequations}\begin{eqnarray}
	\sigma(\mathrm i \Omega_n)&=&-\frac{1}{\Omega_n}\Pi_{xx}(\Omega_n)\\
	\Pi_{xx}(\Omega_n)&=&\frac{1}{\beta L^2}\sum_{i,j} \mathrm{e}^{\mathrm i\Omega_n \tau_{i,j}}\frac{\delta\langle J_{i,i+x}\rangle}{\delta A_{j,j+x}}
	\label{eq:cond_def}
	\end{eqnarray}\end{subequations}
	where $\tau_{i,j}$ is the discrete imaginary time distance between the lattice points $i,j$. 
	We measure the conductivity in units of $Q^2/h$, which amounts to multiplying the conductivity in Eq.\thinspace\eqref{eq:cond_def} by $2\pi$. 
	We study lattices with linear size $L=512$ and we set the microscopic parameter $r=-5.89391$. The critical coupling is then $u_c=7.70285(5)$ as was determined in a previous study \cite{Gazit14}. We made sure that the correlation length, $\xi$, satisfies $\xi<L/2$. 
	In Fig.~\ref{subfig:cond_dis} we give the Matsubara conductivity in the disordered phase for a set of  detuning parameters $\delta u$ in close vicinity to the phase transition. 

	\subsection{Fitting procedure }

\begin{figure}
\center
 \subfigure[]{\label{subfig:cond_dis}\includegraphics[scale=0.55]{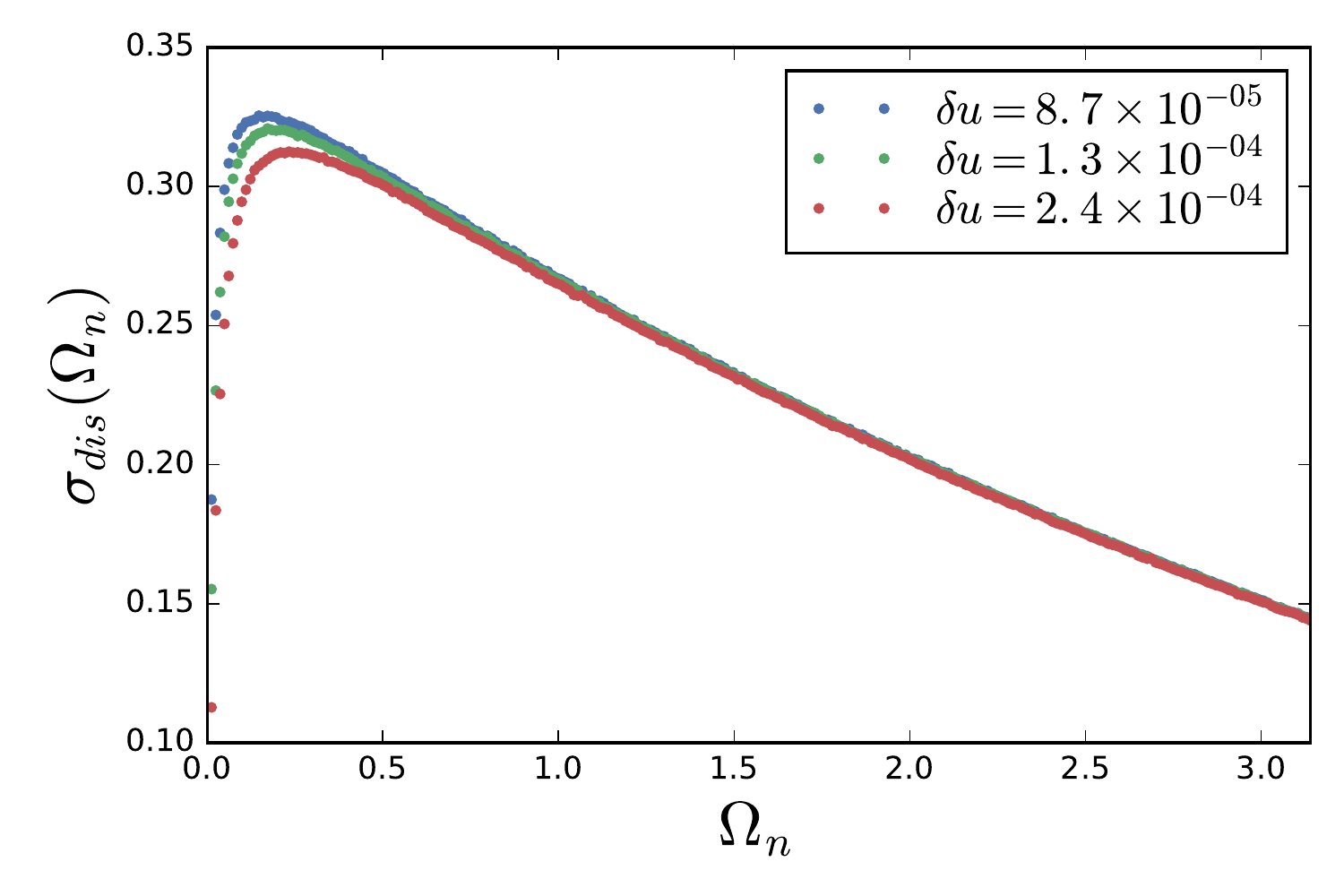}} 
 \subfigure[]{\label{subfig:cond_dis_scale} \includegraphics[scale=0.55]{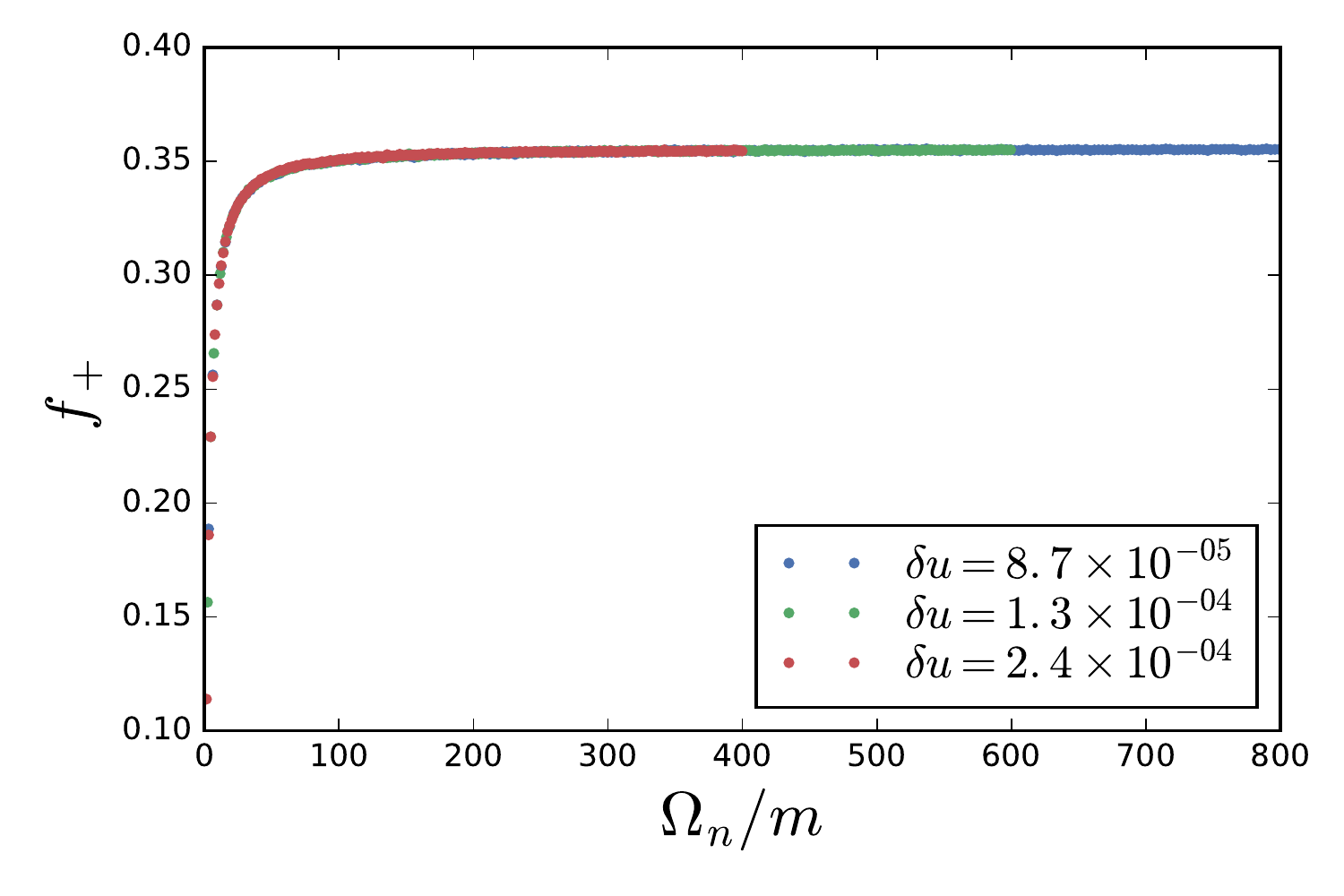}} 
\caption{(a) Conductivity at Matsubara frequencies, $\sigma(\mathrm i\Omega_n) $ in the disordered phase $\delta u>0$. 
(b) Universal scaling function $f_+(\mathrm i\Omega_n/m)$ obtained from Eq.\thinspace(\ref{eq:non_uniscale}) after subtraction the non universal high frequency cutoff corrections to scaling using $\Omega_c/m=100$.  In both panels different curves correspond to difference detuning parameters $\delta u$}
\centering
\end{figure} 

        As discussed in the main text, in two spatial dimensions the conductivity is a universal amplitude and hence near criticality it is expected to follow a scaling form $\sigma(\mathrm i\Omega_n,\delta u)=f_+(\mathrm i\Omega_n/m)$, where $m$ is the single particle gap in the disordered phase. Near criticality the gap vanishes following a power law form $m=m_0(\delta u)^\nu$, with $\nu$ being the correlation length exponent and $m_0$ a non-universal coefficient. 
	 
	 To compute the scaling function from the numerical Monte Carlo data we rescale the Matsubara frequency axis by the single particle gap. We found that at low frequency all curves collapse to a single universal curve, whereas at high frequency we observe  significant deviation from the scaling from. 
		 
	 To understand the origin of these non-universal corrections, we note that lattice discretization inevitably introduces a UV cutoff scale $\Lambda\sim 1/a$ where $a$ is the lattice constant. At large frequency $\Omega_n\gtrsim\Lambda$, the numerical result deviate from the continuum limit as lattice scale effects become sizable. The cutoff scale corrections are expected to be smooth both in $\Omega_n$ and $\delta u$  and we model them using a simple cubic polynomial ansatz 
	\begin{equation}
	\sigma(\mathrm i\Omega_n,\delta u,\Lambda)\approx f_+(\mathrm i \Omega_n/m)+\sum_{l=1}^3 \alpha_l \Omega_n ^l
	\label{eq:non_uniscale}
	\end{equation}
	We further assume that since we study a small range of detuning parameters, the coefficients $\alpha_l$ have a weak dependence on $\delta u$ and we therefore take them to be constants.
	
	Our main task now is to compare the Monte Carlo data with the asymptotic large frequency behavior of the optical conductivity predicted in Eq.~\eqref{eq:asympt}. For $N=2$, the correlation length exponent was estimated in previous high precision Monte Carlo studies to be $\nu=0.6717(3)$ \cite{BurovskiXYnu} such that the power law exponents in Eq.~\eqref{eq:asympt} equal $(d+z-\Delta)/z=1/\nu=1.48987$ and $\Delta/z=3-1/\nu=1.51013$. We see that the two exponents are nearly identical and hence  cannot be resolved within our numerical accuracy. We, therefore, combine them to a single exponent, and consider the following large frequency form for the optical conductivity,
	\begin{equation}
	f_+(x\gg1)\sim\sigma_\infty +C \times\, x^{-1/\nu} 
	\label{eq:asym_scale}
	\end{equation}
	For the infinite  frequency conductivity, we take the high precision bootstrap estimate $\sigma_\infty=0.3554(6)$ \cite{Kos:2015mba}.  This leaves us with four free fitting parameters $C,\alpha_{l=1,2,3}$ that we determine using least square minimization. 
Since the expression in Eq.~(\ref{eq:asym_scale}) is valid only in the high frequency limit, in the numerical fit we only use data points that satisfy $\Omega_c<\Omega_n$. We performed the fit on a range of lower cutoff frequencies $\Omega_c/\gap=50,100,150,200$. 
	
We find that the coefficients $\alpha_l$ are nearly independent of $\Omega_c$ and equal  $\alpha_1\approx -0.1,\alpha_2\approx0.01$ and $\alpha_3\approx-0.001$, working 
in units where the UV cutoff $\Lambda$ (inverse lattice spacing) has been set to unity. 
We subtract the cutoff scale corrections and plot the universal scaling function $f_+$ in Fig.~\ref{subfig:cond_dis_scale}. Our estimate for the power-law coefficient is $C=-5.0(5)$. The quoted numerical error is 
dominated by variations with respect to the lower cutoff frequency $\Omega_c$. For curve plotted in Fig.~\ref{fig:qmc} in the main text we used $\Omega_c/m=100$, for which $C=-4.83$ and the reduced goodness of fit equals $\chi^2=0.93$.

	As a final remark, we wish to emphasize that although our results are consistent with the predicted scaling form, our numerical analysis involves subtraction of a non-universal background signal that is relatively large compared to the high frequency component of the universal scaling function. As a consequence, we did not manage to extract an independent estimate for the predicted power law exponents. Improving the numerical scheme for eliminating the large frequency corrections to scaling is an interesting line of research that we intend to study in the future.

\bibliography{detuningbib}
\end{document}